\documentclass[useAMS,usenatbib]{mn2e}
\usepackage{natbib}
\usepackage{graphicx}
\usepackage{float}

\usepackage{subfigure}
\usepackage{amsmath}
\usepackage{amssymb}
\usepackage{cite}
\title[Star forming region IRAS 20286+4105]{Radio and infrared study of
the star forming region IRAS 20286+4105}
\author[Varsha R et al.]{Varsha R$^{1}$\thanks{Present Affiliation and E-mail: Institute of Physics and Astronomy, University of Potsdam; varsha@astro.physik.uni-potsdam.de}, S. R. Das$^{1}$, A. Tej$^{1}$, S. Vig$^{1}$, S. K. Ghosh$^{2,3}$, D. K. Ojha$ ^{3} $\\
$^{1}$Indian Institute Of Space Science And Technology, Trivandrum, India\\
$ ^{2}$National Centre for Radio Astrophysics (NCRA-TIFR), Pune, India \\
$ ^{3}$Tata Institute of Fundamental Research (TIFR), Mumbai, India
}

\begin{document}

\date{}

\pagerange{\pageref{firstpage}--\pageref{lastpage}} \pubyear{2015}

\maketitle

\begin{center}
\begin{abstract}
A multi-wavelength investigation of the star forming complex IRAS 20286+4105, located
in the Cygnus-X region, is presented here. Near-infrared K-band data is used to revisit the 
cluster / stellar group identified in previous studies. 
The radio continuum observations, at 610 and 1280~MHz show the presence of a
HII region possibly powered by a star of spectral type B0 - B0.5. The cometary morphology of
the ionized region is explained by invoking the bow-shock model
where the likely association with a nearby supernova remnant is also explored. 
A compact radio knot with non-thermal spectral index is detected towards
the centre of the cloud. 
Mid-infrared data from the {\it Spitzer} Legacy Survey of the Cygnus-X region
show the presence of six Class I YSOs inside the cloud. 
Thermal dust emission in this complex is modelled using {\it Herschel} 
far-infrared data to generate dust temperature and column density 
maps. {\it Herschel} images also show the presence of two clumps in this region, 
the masses of which are estimated to be $\sim 175\,{\rm M}_{\odot}$ and 30$\,{\rm 
M}_{\odot}$. The mass-radius relation and the surface
density of the clumps do not qualify them as massive star forming sites. An overall
picture of a runaway star ionizing the cloud and a triggered population of intermediate-mass, 
Class I sources located toward the cloud centre emerges from this multiwavelength study.
Variation in the dust emissivity spectral index is shown to exist in this region and is seen to 
have an inverse relation with the dust temperature.
  
\end{abstract}

\end{center}

\section{Introduction}
\label{intro}

IRAS 20286+4105 (Mol 126; G79.873+1.179; $\alpha_{2000}$ = $\rm 20^h30^m29.9^s$, $\delta_{2000}$ = +$41^{\circ}$15\arcmin51\arcsec) is a star 
forming complex in the Cygnus - X region with a far-infrared (FIR) luminosity of 
5.3 $ \times 10^{3} $ L$ _{\odot} $ \citep{2011ApJ...727..114R}.
The distance estimates for this region vary from $\sim$ 1~kpc \citep{{2001A&A...375..539C}, 
{1989ApJ...345L..47O}} to $\sim$ 4~kpc \citep{1991A&A...246..249P}. In this paper we adopt 
a recent distance estimate of 1.61~kpc, obtained from maser astrometry 
\citep{2013ApJ...769...15X}. This is consistent with the adopted values in recent 
studies of the Cygnus region \citep{{2006A&A...458..855S},{2011ApJ...727..114R}}.  
Based on its luminosity and IRAS colours, IRAS 20286+4105
has been proposed as a candidate high-mass protostellar object \citep{1996A&A...308..573M}. 
From $ ^{12} $CO J = 2-1 observations, 
\citet{1989ApJ...345L..47O} were able to resolve IRAS 20286+4105 as a compact CO hotspot
and showed its association with an elliptical thin luminous ring seen in the optical wavelengths.
Based on 2MASS data, a resolved cluster/stellar group is also reported to be associated with 
this region \citep{{2001A&A...376..434D}, {2006A&A...449.1033K}}. This 
complex has been the target of a couple of radio continuum observations
\citep{{1984A&AS...58..291W}, {1991A&A...241..551W}, {1991AJ....101.1435M}}. In the
FIR, this region has been studied in detail by \citet{2003A&A...398..589V} at 150 and 
210~$\rm \mu m$ using the balloon-borne TIFR telescope. IRAS 20286+4105 also forms part
of the Balloon-borne Large Aperture Submillimeter Telescope
(BLAST) survey of the Cygnus region at 250, 350 and 500~$\rm \mu m$ \citep{2011ApJ...727..114R}.
 Based on the near-infrared (NIR) $\rm H_{2}$ line maps, \citet{2010MNRAS.404..661V} have
identified two outflows, the positions of  which are consistent with the 
CO outflow results of \citet{2005ApJ...625..864Z}. 

IRAS 20286+4105 has also been the target of many 
surveys to detect masers. Water, ammonia
and Class I Methanol masers have been detected towards this region \citep{{1991A&A...246..249P},
{1994A&AS..103..541B}, {1996A&A...311..971C}, {1996A&A...308..573M}, 
{2011MNRAS.418.1689U}, {2013ApJ...763....2G}}.
However, the search for 
Class II methanol and hydroxyl masers yielded negative results \citep{{1995A&AS..110...81V},
{2007A&A...465..865E}}. Several molecules like  CS, HCO$ ^{+} $, NH$_3$, N$ _{2} $H$ ^{+} $, CH$_{3}$CCH 
have also been observed towards this region \citep{{1996A&AS..115...81B},{2002ARep...46..551A}, 
{2013ApJS..209....2S},{2014ApJ...790...84L}}.   

In this paper, we present an in-depth observational study of the IRAS 20286+4105 star forming complex from infrared through radio wavelengths.
Low frequency radio continuum 
observations using the Giant Metrewave Radio Telescope (GMRT) enables us to probe the associated
ionized emission at various spatial scales. UKIRT Infrared Deep Sky Survey (UKIDSS) NIR 
and {\it Spitzer} Legacy Survey of the Cygnus - X complex mid-infrared (MIR) data allows us to study in 
detail the related stellar population. FIR data from the 
{\it Herschel} infrared Galactic Plane Survey (Hi-Gal) allows us to understand the dust environment 
and study the star forming activities therein. 
In Section \ref{obvs.sect}, we discuss these observations and related
data reduction procedures used. This section also gives the details of the various datasets retrieved
from archives and used in the present study. Section \ref{results} gives a comprehensive discussion
on the results obtained and Section \ref{conclusion} summarizes the results.

\section{Observations, data reduction and archival data}
\label{obvs.sect}

\subsection{Radio continuum observations}
\qquad The ionized gas emission associated with IRAS 20286+4105 is probed using radio continuum 
interferometric mapping with the GMRT. The basic 
structure of GMRT is a `Y'- shaped hybrid configuration of 30 antennae (of 45~m diameter each), 
in which six antennae
each are placed along the three arms of length $\sim$ 14~km each. These provide high angular 
resolution (with longest baseline $\sim$ 25~km). The other twelve antennae 
have a central compact arrangement within a $\rm 1\times1$ $\rm km^{2}$ area sensitive to detection of
large scale diffuse emission. Technical details regarding GMRT can be found in \citet{1991ASPC...19..376S}. 

Continuum observations were carried out at 1280 and 610~MHz. Radio sources 3C48 and 3C147 (primary flux 
calibrators), 2022+422 and 2052+365 (phase calibrators) were observed for 
flux and phase calibration of the measured visibilities. 
Data reduction is performed using the Astronomical Image Processing System (AIPS). Removing bad 
data (owing to dead antennas, bad baselines, interference, spikes, phase variations, etc) is crucial
for obtaining a good map. This is accomplished iteratively using editing 
and flagging tasks UVPLT, VPLOT, TVFLG and UVFLG in combination with calibration tasks CALIB, SETJY, GETJY, and CLCAL. Next, task SPLIT is used to separate out the
source data. Subsequent to this, facets are generated using the task SETFC. The intensity map
is then obtained using the Fourier inversion and cleaning algorithm task IMAGR. Several iterations
of self calibration are performed in order to minimize amplitude and phase errors. Following this,
the facets are combined using the task FLATN and primary beam correction is applied using the task PBCOR.
While observing close to the Galactic plane, the contribution of diffuse background emission to the
system temperature in the frequency domain of our observations is appreciable and hence cannot be neglected. The generated maps are corrected based on the estimation of the
sky temperature from the 408~MHz all sky survey of \citet{1982A&AS...47....1H} and using a spectral index of 
-2.7 \citep{2003A&A...403.1031H} for Galactic diffuse emission. A detailed description
of the system temperature correction for GMRT data can be found in \citet{2015MNRAS.451...59M}. Final images are obtained by using scaling factors of 1.7 and 1.1 at 610 and 1280~MHz, respectively. 

\subsection{Near-infrared data from HCT-TIRSPEC}
\label{hct_data}

NIR narrow-band imaging was carried out using TIFR Near Infrared Spectrometer and Imager (TIRSPEC) mounted on
the 2-m Himalayan {\it Chandra} Telescope (HCT). The instrument consists of a~1024~$\times$~1024 pixel HgCdTe Teledyne 
Hawaii-1 PACE array detector. The imaging mode of the instrument gives a plate scale of 0.3\arcsec~per pixel, 
resulting in a~307$\arcsec\times$~307\arcsec~field of view. Detailed description of TIRSPEC can be found in  
\citet{2012ASInC...4..191O} and \citet{2014JAI.....350006N}.
Imaging observations in the $\rm Br\gamma$ and K-Cont narrow-band filters were carried out in the five-point 
dithered mode with several exposures at each dithered position. Sky frames were also obtained in the same mode.  
Standard methods were adopted to obtain flat frames. Details of NIR observations are listed in Table \ref{tab_tirspec}. 

\begin{table*}
\begin{center}
\caption{Details of narrow-band imaging using TIRSPEC.}
\begin{tabular}{|c|c|c|c|c|c|}
\hline
Observation date & Filter & Wavelength  & Bandwidth & Exposure time/frame  & Total integration time  \\
& & ($\rm \mu m$) & (\%)&(secs)&(secs) \\
\hline
02 June 2014 & $\rm Br\gamma$ & 2.166&0.98 & 40 & 1000 \\
02 June 2014 & K-Cont & 2.273&1.73 & 40 & 1000 \\
\hline   
\label{tab_tirspec}
\end{tabular}
\end{center}
\end{table*} 

The initial process of dark subtraction is done with the inbuilt instrument software. 
Subsequent to this a semi automated pipeline\footnote{https://github.com/indiajoe/TIRSPEC} is used 
for further image reduction. Flagging of bad frames is done in an interactive mode. Standard 
procedures of sky subtraction, flat fielding and bad pixel masking is carried out. Finally, the 
images taken at different dither positions are aligned (with reference to a few isolated stars) 
and combined to generate final images in the two narrow band filters. In order to probe the ionized emission in the
$\rm Br\gamma$ line, aligned and scaled K-Cont image is subtracted from the $\rm Br\gamma$ image. 
$\rm Br\gamma$ and K-Cont are both narrow-band 
filters with no significant difference in the FWHM of the stellar images, hence we have not 
done PSF matching. However, scaling is necessary to account for the slight difference in filter 
bandwidths and possible variation in seeing conditions. The scaling factor is estimated from isolated bright stars in the images. The continuum-subtracted image is further binned to improve the signal-to-noise ratio. 

\subsection{Available datasets from various archives}
\subsubsection{4.89 GHz radio map from VLA archive}

The 4.89 GHz (C-band) radio continuum map for this region has been obtained from NRAO VLA archive\footnote{This NVAS image was produced
as part of the NRAO VLA Archive Survey.}. This complex was observed in April 1988 with 27 VLA antennae in the C/C 
configuration and the retrieved image has been reduced and calibrated using the VLA pipeline in March 2009. 
The retrieved map has a beam size of 4.07\arcsec$\times$~4.01\arcsec, field of view of 4.56\arcmin~and {\it rms} level $ \sim $ 100~$\mu$Jy/beam. 
Though the date of observation is not explicitly mentioned in their paper, this data is possibly the one which is
presented in \citet{1991AJ....101.1435M}. We have used this data alongwith the data obtained from GMRT maps to study the ionized emission associated with IRAS 20286+4105.

\subsubsection{Near-infrared data from UKIDSS}
\label{ukidss_data} 
In this work we use K-band (2.20~$\rm \mu m$) data from the UKIDSS 8PLUS Galactic Plane Survey (GPS) which has a  resolution of~$\sim$~1$\arcsec$ and 5~$\sigma$ limiting magnitude of K = 18.1 mag \citep{2006MNRAS.372.1227D}. As discussed in \citet{2008MNRAS.391..136L}, 
in order to get an estimate of the star count in our field of interest, we 
filter the sources classified as `noise' and retain those with {\it mergedClass} values of -1 or -2 that denotes a star or probable star. 
In addition, we use the attribute {\it PriOrSec} to account
for the duplicate catalog entry in the overlapping region of the WFCAM tiles. The retrieved data 
set goes up to the faint limit of 19.2 mag.

\subsubsection{Mid-infrared data from Spitzer Space Telescope}
MIR images and photometric data in the four IRAC bands and MIPS~24~$\rm \mu m$ band around our region of 
interest have been retrieved from the archives of the {\it Spitzer} Legacy Survey of the Cygnus - X complex. This project aims at surveying a 24 square degrees region with the four IRAC bands and two MIPS bands \citep{2007sptz.prop40184H}. The images have a resolution of 1.6\arcsec, 1.6\arcsec, 1.8\arcsec, 1.9\arcsec, 6\arcsec~ in 3.6, 4.5, 5.8, 8.0, 24~$\rm \mu m$ bands, respectively.

\subsubsection{Far-infrared data from Herschel Space Observatory}
The FIR maps of the region around IRAS 20286+4105 in the wavelength range 70 - 500~$\rm \mu m$ are obtained from 
the archive of the Hi-Gal survey. Hi-Gal is a key 
project of the {\it Herschel} Space Observatory for mapping the inner Galactic plane, covering $ |l|\leq 60^{\circ} $ 
and $|b|\leq 1^{\circ}$, for probing various phases of massive star formation \citep{2010A&A...518L.100M}. 
The Level - 3 mosaic products of SPIRE maps at wavelengths 250, 350, and 500~$\rm \mu m $ with SPIRE calibration 
level 12.5 and PACS maps at 70 and 160~$\rm \mu m $ with calibration level 2.5 have been retrieved. These retrieved images are scanned in parallel 
mode and calibrated using  {\it Herschel} Interactive Processing
Environment (HIPE). The resolutions of the images are 5.9\arcsec, 11.6\arcsec, 18.5\arcsec, 25.3\arcsec, and 36.9\arcsec~at 70, 160, 250, 350, and 500~$\rm \mu m$, respectively. The maps have different pixel
sizes ranging from 3.2\arcsec to 14\arcsec. 

\subsubsection{H$ \alpha $ image from IPHAS}
$\rm H\alpha$ ($\rm \lambda=656.83~nm;~\Delta \lambda=9.6~nm$) and $\rm r^{\prime}$-band ($\rm \lambda=625.41~nm;~ 
\Delta \lambda=135.0~nm$) observations of the region around IRAS 20286+4105 have been
carried out at the Isaac Newton Telescope (INT) in La Palma. The images were
retrieved from the archive of INT 
Photometric H$\alpha $ Survey (IPHAS) \citep{2008MNRAS.388...89G}. This is an imaging survey of the Northern Milky Way in visible 
light (H$\alpha $, r$^{\prime}$, i bands) down to $ > $ 20th magnitude in the r$^{\prime}$-band. The resolution of the images are $\sim$~1.3\arcsec. 
Using the r$^{\prime}$-band image and following the method discussed in Section \ref{hct_data}, we obtain the continuum-subtracted $\rm H\alpha$ line image.

\section{Results and discussion}
\label{results}

\subsection{Revisiting the cluster associated with IRAS 20286+4105}
\label{cluster}

\begin{figure*}
\includegraphics[scale=0.36,bb=15 20 600 650,clip]{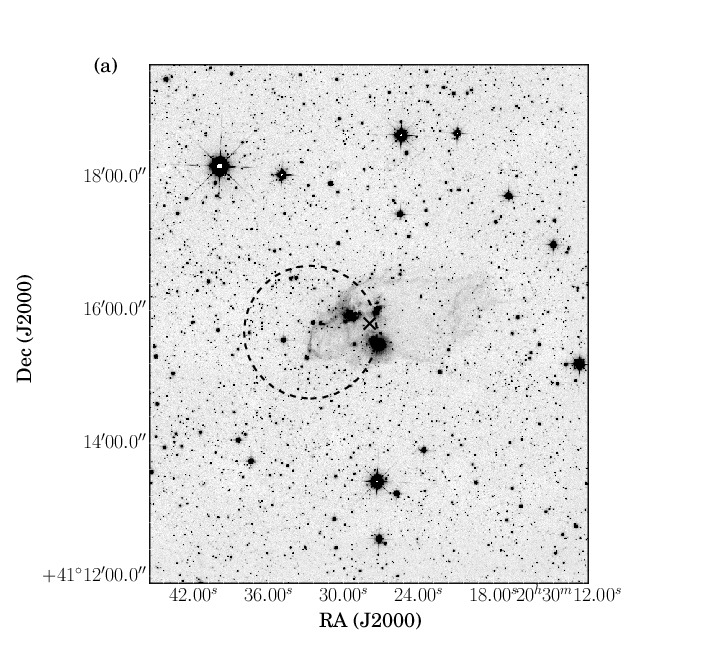} 
\includegraphics[scale=0.36,bb=15 20 600 650,clip]{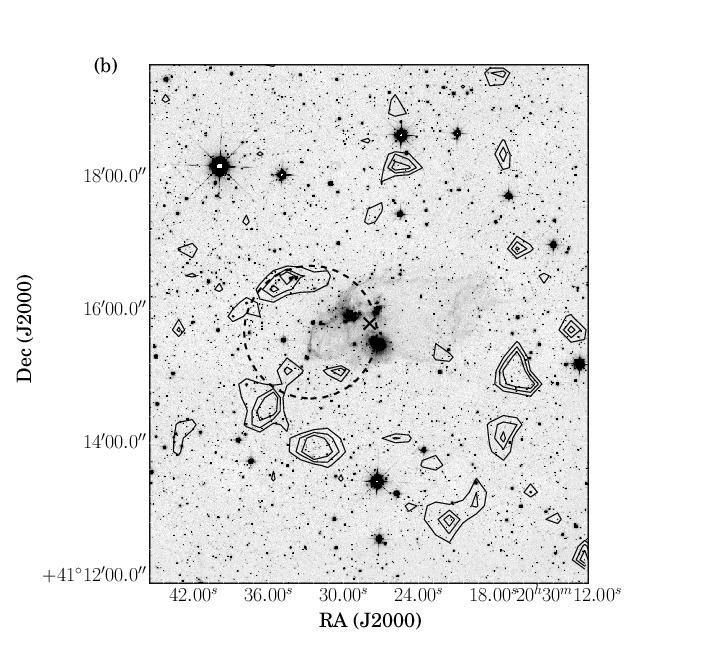}
\caption{Stellar surface density contours using UKIDSS K-band data overlaid on the 
K-band UKIDSS image of the region associated with IRAS 20286+4105 - (a) results obtained using the star count method where no contours are seen above the mode+2$\sigma$ level (b) 
results obtained with NN algorithm. Contour levels are from mode+2$\sigma$ in steps of 1$\sigma$. The $\times$ 
sign represents the position of the IRAS point source. The circle shows the expected position
and extent of the cluster as estimated by \citet{2006A&A...449.1033K}.}
\label{sc_nn}
\end{figure*}

A sparsely populated cluster with 7 `true' cluster members has been shown to be associated with IRAS 20286+4105 
\citep{2006A&A...449.1033K}. These authors have used the 2MASS K-band data and the star count method to detect this 
cluster. With
the availability of higher sensitivity and better resolution UKIDSS data, we aim at studying the cluster in
detail. 
As mentioned in Section \ref{ukidss_data}, we use the K-band photometric data from the pipeline processed UKIDSS 8PLUS GPS catalog.
The first step towards this is to identify the cluster before proceeding to study the physical 
properties of the stellar population associated with it. 
We choose the K-band which gives the advantage 
of sampling the more embedded members of the cluster. We select an area within 300\arcsec radius centered on the position of the IRAS point source. 

Initially, we adopt the star count method outlined in \citet{2011AN....332..172S}. The entire field is subdivided into 
rectilinear grids of overlapping squares. For maintaining Nyquist special sampling 
interval, the separation between the squares is made to be half the length of the squares  
\citep{1995AJ....109.1682L}. This method then involves determining the number of stars in each 
individual square grid. Each square grid acts as a single point 
for producing the surface density map. The square grids with star counts greater than some 
significant threshold ($ \sim $ 2 - 5$ \sigma $) above the background are regions
of density enhancements and hence are considered as potential cluster locations \citep{2011AN....332..172S}. 
In our case, the mode (0.019 stars $\rm arcsec^{-2}$ or 312 stars $\rm pc^{-2}$) of the grids 
is considered to be the background and the noise/fluctuation in them defines $ \sigma $ which
is estimated to be 0.005 stars $\rm arcsec^{-2}$ or 82 stars $\rm pc^{-2}$. We assume mode+2$\sigma$ as the
threshold for identifying density enhancement centers. Deciding the grid size is a crucial factor. We vary the grid size from 
60$\arcsec$ to 110$\arcsec$ and find 80$\arcsec$ to be optimal. This is also the value adopted by 
\citet{2006A&A...449.1033K}. Figure \ref{sc_nn} shows the K-band image and
the stellar density contours above the mode+2$\sigma$ threshold. 
As is clearly evident from the figure, there are no stellar density contours
above the mode+2$\sigma$ level associated with the IRAS source. In fact, the entire
field does not show any density enhancement above the defined threshold.
The dotted circle roughly shows the position and extent of the cluster as estimated by \citet{2006A&A...449.1033K}. 

\begin{figure*}
\includegraphics[scale=0.33,bb=145 60 600 590,clip]{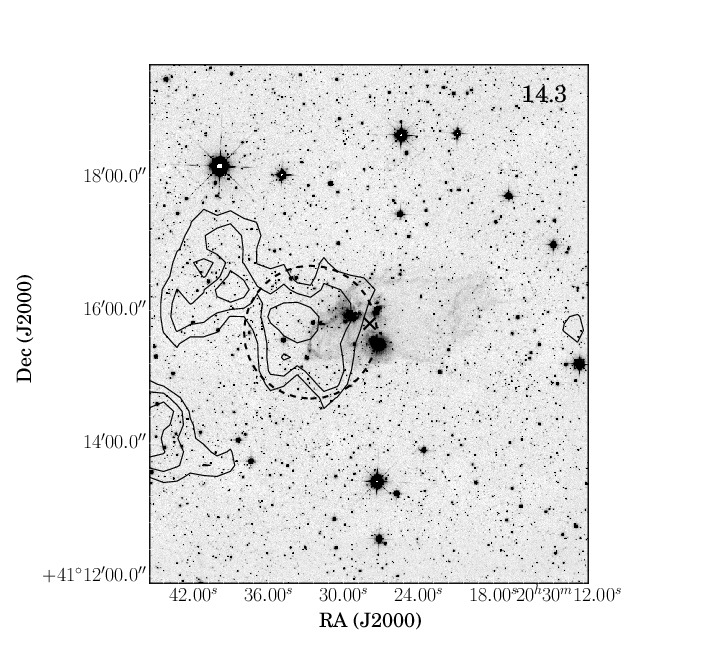} 
\includegraphics[scale=0.33,bb=145 60 600 590,clip]{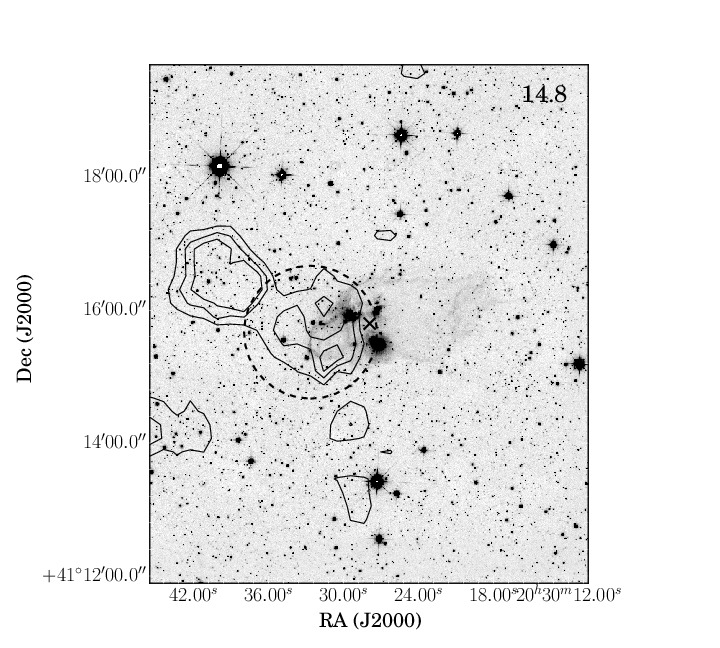}
\includegraphics[scale=0.33,bb=145 60 600 590,clip]{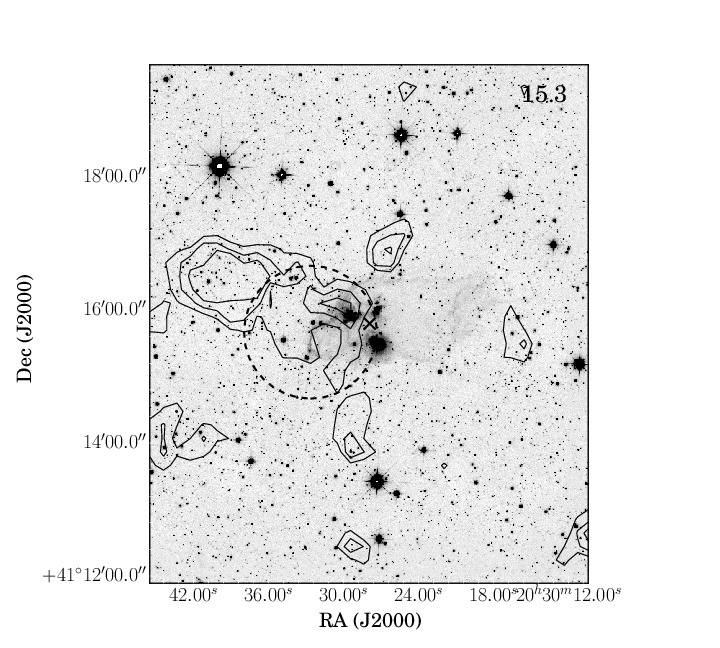} 
\includegraphics[scale=0.33,bb=145 60 600 590,clip]{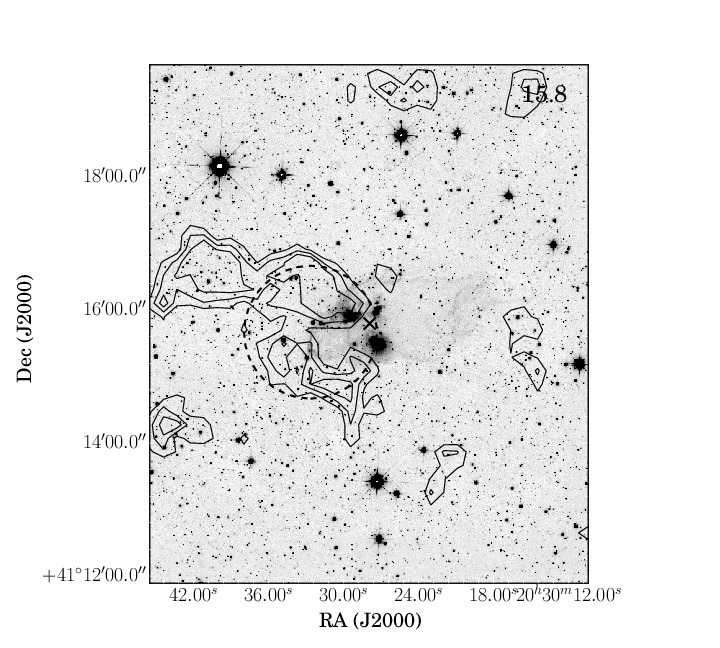}
\includegraphics[scale=0.33,bb=145 60 600 590,clip]{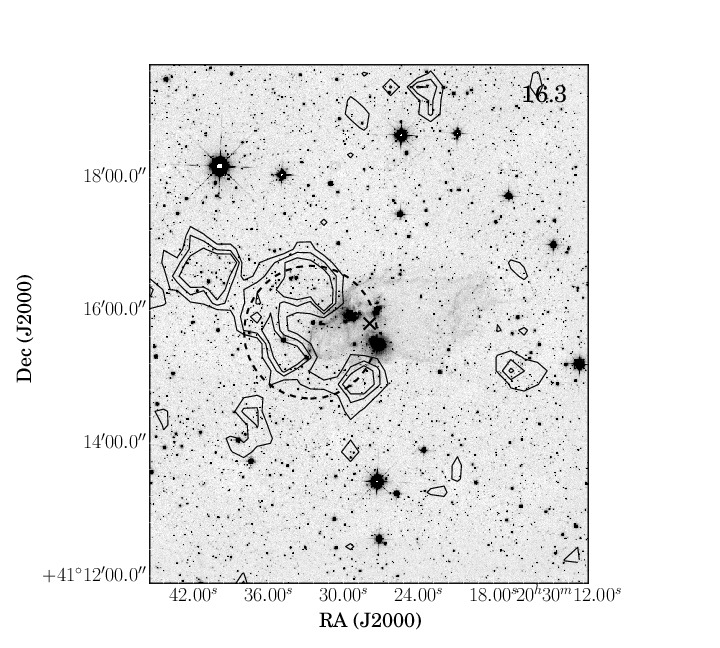} 
\includegraphics[scale=0.33,bb=145 60 600 590,clip]{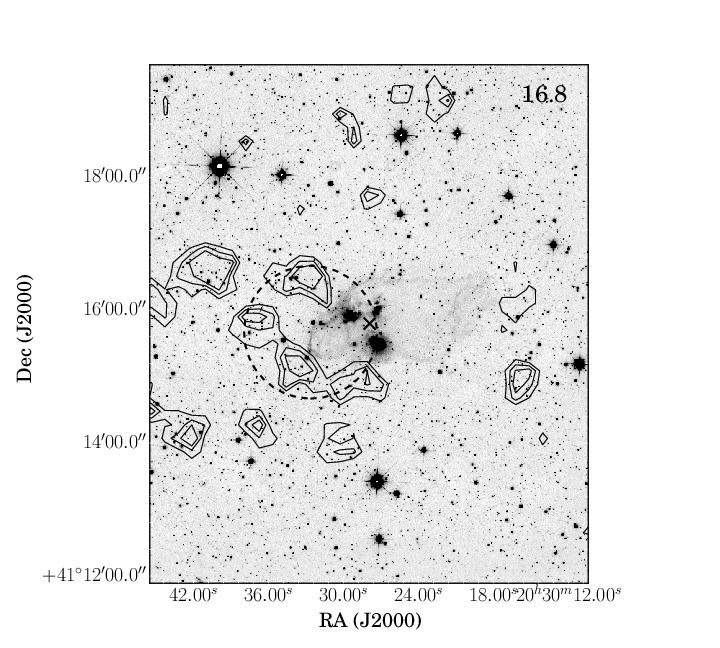}
\includegraphics[scale=0.33,bb=145 60 600 590,clip]{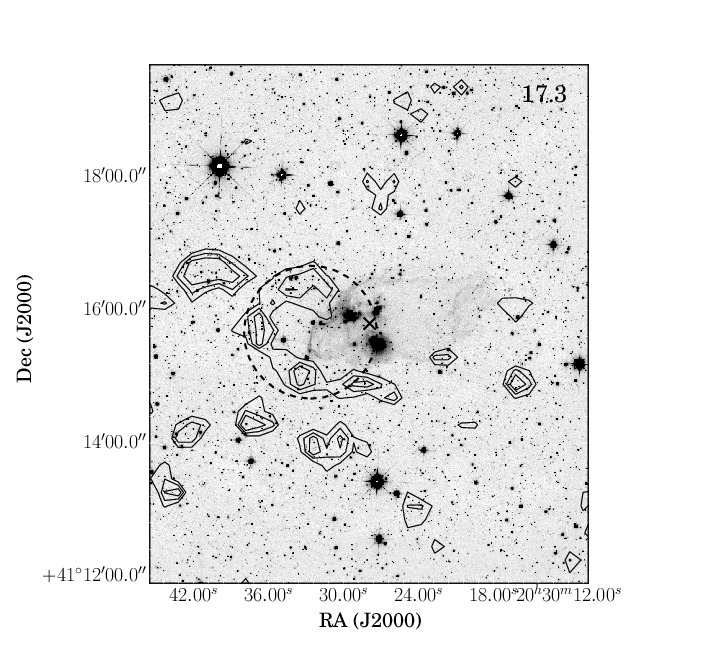} 
\includegraphics[scale=0.33,bb=145 60 600 590,clip]{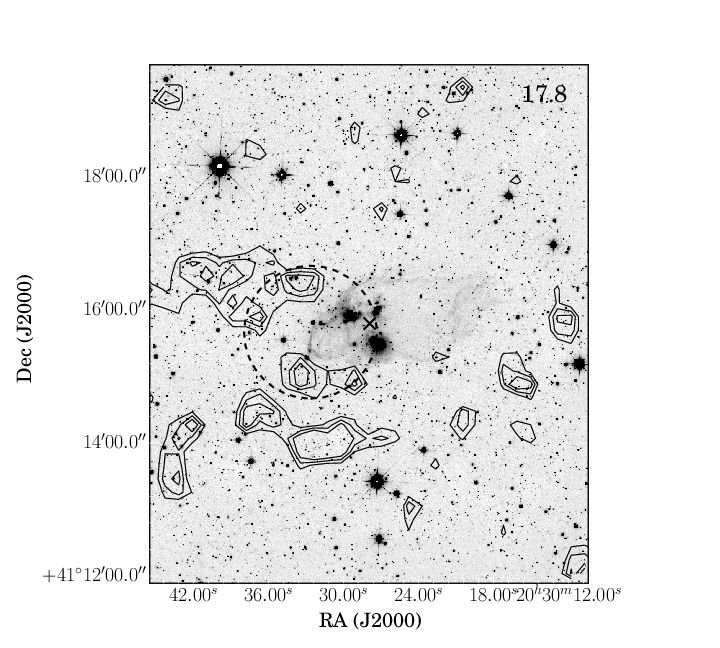} 
\includegraphics[scale=0.33,bb=145 60 600 590,clip]{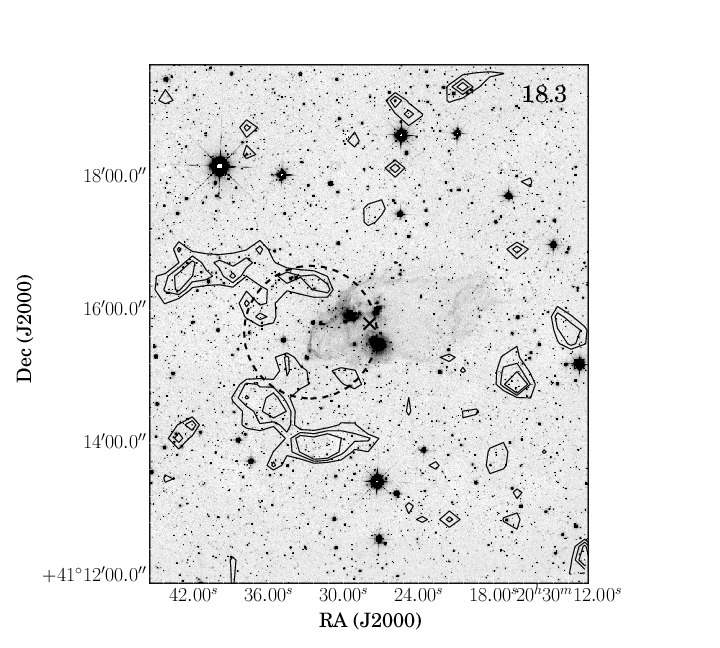} 
\caption{Stellar surface density contours using the NN algorithm using UKDISS data for different magnitude
limits. The magnitude limit starts from the 2MASS limit of 14.3 mag and is increased in steps of 0.5 mag till the
UKIDSS limit of 18.1 mag is reached. Contour levels are from mode+2$\sigma$ in steps of 1$\sigma$ overlaid on the
UKIDSS K-band image. The $\times$ sign represents the position of the IRAS point source. The field of view of the images is same as in Figure \ref{sc_nn}. The circle shows the expected position
and extent of the cluster as estimated by \citet{2006A&A...449.1033K}.}
\label{mlim}
\end{figure*}

In order to confirm the non-detection of the cluster, we use another 
cluster identification technique namely the nearest neighbour (NN) algorithm on the same dataset. 
Following the procedure outlined in \citet{2011AN....332..172S} and \citet{1985ApJ...298...80C}, the $j^{\rm th}$ nearest neighbour density is defined as

\begin{equation}
 \rho_{j} = \dfrac{j-1}{S(r_{j})} 
\end{equation}
where $r_{j} $ is the distance of a star
to its $j^{\rm th}$ nearest neighbor, $S(r_j)$ is the surface area within $r_{j} $. NN density method 
depends only on the value of $j$. As given in \citet{2008MNRAS.389.1209S}, $j$ = 20 is  an 
optimum choice to detect clusters with 10 to 1500 members. We experimented on a range of values for $j$ and also 
found $j$ = 20 to be optimal. Low values of $j$ are sensitive to very small scale statistical fluctuations and higher values miss out the real small scale 
enhancements. The mode of the density values (0.026 stars $\rm arcsec^{-2} $ or 427 stars $\rm pc^{-2}$) gives 
the background and the fluctuation, $\sigma $ (0.004 per $\rm arcsec^{-2} $ or 66 stars $\rm pc^{-2}$) is 
the noise level. The results remain the same
with this method as well and shown in Figure \ref{sc_nn}. We do not see any stellar density 
enhancement at the expected location of the previously detected cluster. 
However, few small-scale clusterings are seen to be present towards the periphery
of the expected location of the cluster. Similar small-scale features are also seen elsewhere
in the field.

In this paper we have used higher sensitivity and better resolution UKIDSS data in comparison to the
2MASS data used by \citet{2006A&A...449.1033K}.
The UKIDSS has a 5$ \sigma $ limiting magnitude of 18.1 mag in the K-band 
\citep{{2006MNRAS.372.1227D}, {2007MNRAS.379.1599L}}, whereas the K-band magnitude
limit (10$ \sigma $) of 2MASS is 14.3 mag \citep{2006AJ....131.1163S}.
This prompted us to study the effect of sensitivity on cluster detection
using the conventional techniques. 
This is achieved by repeating the above 
exercise with magnitude cuts starting from the 2MASS limit of 14.3 mag and proceeding in steps of 0.5 mag till we reach the 5$ \sigma $ UKIDSS limit. Figure \ref{mlim} shows the result at each magnitude limit. As seen in the figure, the stellar density enhancement is clearly evident at the 2MASS limit of
14.3 mag and is fairly consistent with the result obtained by \citet{2006A&A...449.1033K} (see their Fig. 1). In comparison to the results obtained by \citet{2006A&A...449.1033K}, 
the cluster seems to be extended in the north-east direction. The signature of the cluster is clearly visible upto the magnitude
cut-off of 16.3 mag. However, beyond this limit, the surface density contours appear to be fragmented into small scale enhancements mostly towards the eastern side of the expected location of the cluster.  

This probably can be understood from the following consideration. 
As the magnitude limit is increased, the data set probes fainter sources which implies 
increase in source density. This increase includes both genuine cluster members as well as 
background contamination in the line of sight. In this case, the background increases 
by more than a factor of 10 from $\sim$ 25 stars $\rm pc^{-2}$ at 14.3 mag cut to $\sim$ 330 
stars $\rm pc^{-2}$ at 18.3 mag which is close to the 5$ \sigma $ UKIDSS limit. Given the fact that the
cluster associated with IRAS 20286+4105 is a sparsely populated one, it is likely that
the overwhelming background suppresses the cluster stellar density
enhancement as we proceed towards fainter limits.
The small scale features can be taken to be mostly due to fluctuations in the background. 
Another reason for these small-scale 
fluctuations could be the apparent over density of sources towards the east of the
IRAS 20286+4105 cloud which is shown in later sections to harbour two dense clumps
with low stellar content.
Given the non-detection of the cluster using the new UKIDSS data, we could not 
proceed further towards a detailed study. In order to gain a better insight into
the results obtained, in a forthcoming work (in preparation), we aim at
using UKIDSS data for investigating this effect of sensitivity on cluster detection for a large sample of clusters already detected with 2MASS data.

\subsection{Stellar population around IRAS 20286+4105}
\label{population}

\subsubsection{Identification of YSOs}
\label{yso}

IRAS 20286+4105 is a star forming complex in which previous studies have indicated the presence of young stellar objects (YSOs) \citep{2010MNRAS.404..661V}. In this section we study the nature of the stellar population around IRAS 20286+4105 and identify candidate YSOs in this region. We use IRAC photometric data for this study.
MIR colours are efficient tools to identify young stars, especially to distinguish between 
stars with disks and envelopes \citep{{2004ApJS..154..363A}, {2004ApJS..154..367M}, 
{2007ApJ...669..327S}, {2008ApJ...674..336G}}.  In comparison with the NIR (e.g. JHK bands) emission 
from young stars, the MIR IRAC bands sample a significant fraction of emission from circumstellar 
material over the stellar photosphere. The IRAC images of the region associated with IRAS 20286+4105 display
an elliptical shaped cloud. To ensure that we include the entire cloud and also a significant portion of the 
surrounding region, we consider a region of radius 150$\arcsec$ centered
on the position of the IRAS point source. We detect 100 sources within this region that have good quality data in all four IRAC bands (3.6, 4.5, 5.8 and 8 $\rm \mu m $). We adopt the following four schemes to identify candidate YSOs.
It is worth mentioning here that we are considering only those sources detected in
all four IRAC bands hence probing only a sub-sample of YSOs. 
\begin{figure*}
\centering
\includegraphics[scale=.4]{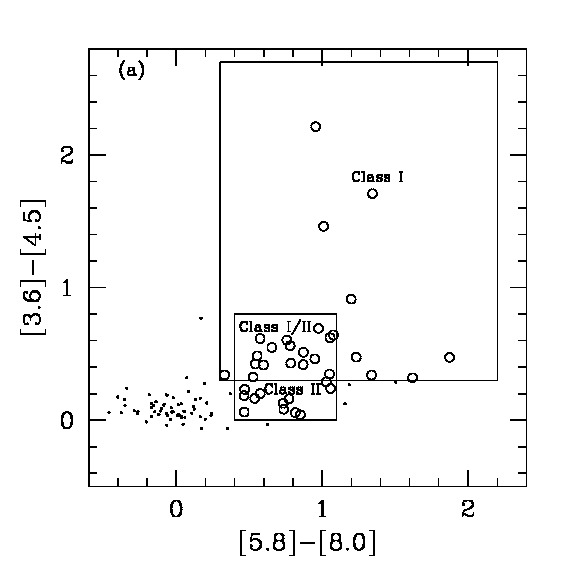}
\includegraphics[scale=.4]{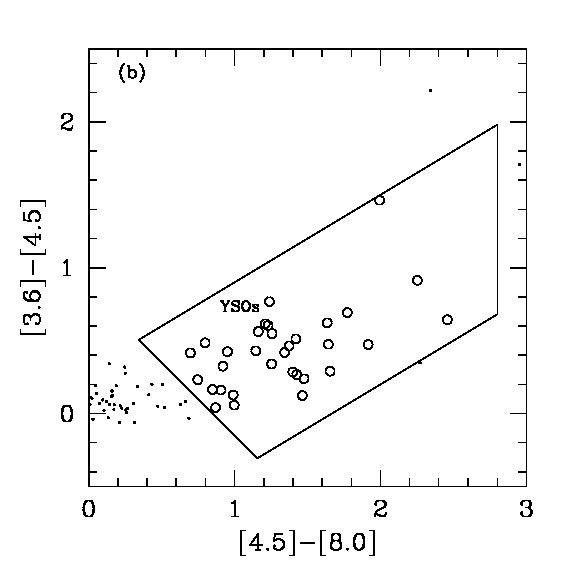}
\includegraphics[scale=.4]{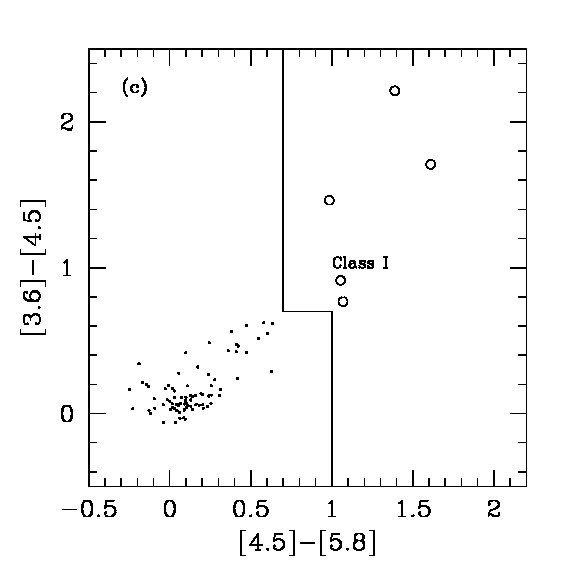}
\includegraphics[scale=.4]{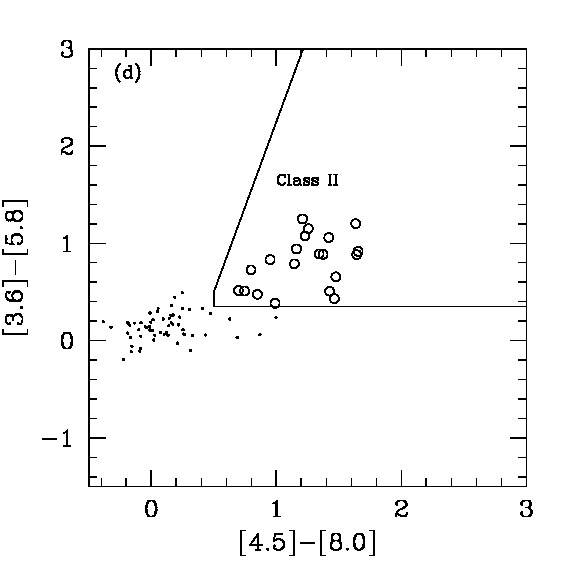}
\caption{ Colour-colour diagrams for the schemes discussed. (a) \citet{2004ApJS..154..363A}  (b) \citet{2007ApJ...669..327S} (c) \citet{2008ApJ...674..336G} Class I and (d) \citet{2008ApJ...674..336G} Class II. The open circles represent the identified YSOs.}
\label{spz_class}
\end{figure*}

\begin{enumerate}
\item  \citet{1987IAUS..115....1L} define the IRAC spectral index to be
\begin{equation}
 \alpha_{\rm IRAC} = \dfrac{{\rm log}(\lambda F_{\lambda})}{{\rm log}(\lambda)}
 \end{equation} 
which is then used to classify YSOs. Using linear regression to the observed IRAC
fluxes, we estimate the spectral index and classify the sources based on the criteria
outlined in \citet{2008ApJ...682..445C}. Class I sources ($ \alpha >$  0) are  
embedded sources with circumstellar disks and envelopes; Class II, or classical T Tauri stars 
(-2 $ < \alpha < $ 0) have significant circumstellar disks, strong emission lines and 
substantial IR or UV excesses; Class III, or weak emission T Tauri stars ($ \alpha <$ -2) have 
weak or no emission lines and negligible excesses. Following the above criteria, we identify 10 Class I and 37 Class II sources. 

\item IRAC [3.6] -- [4.5] vs [5.8] -- [8.0] color-color (CC) diagram is also useful
 for classifying proto-stellar objects into their various evolutionary stages, 
such as Class I, Class II and Class III \citep{2004ApJS..154..363A}. 
Class I and Class II sources are identified based on their location in the CC plot.
We identify 9 Class I, 16 Class I/II and 12 Class II candidate YSOs using boxes to demarcate 
the regions occupied by Class I and Class II sources as discussed in \citet{2007A&A...463..175V}.

\item The third scheme is based on the method proposed by \citet{2007ApJ...669..327S}. They use the [3.6] -- [4.5] and
[4.5] -- [8.0] colours to identify likely YSOs. Their set of colour criteria, which include the 
removal of galaxies and sources with contamination from polycyclic aromatic hydrocarbon emission, do not differentiate
 between Class I and II sources. Following this we identify 32 YSOs in our region. 

\item \citet{2008ApJ...674..336G} also use [3.6] -- [4.5] vs [4.5] -- [5.8] CC diagram for identifying protostellar candidate (Class I) sources and [3.6] -- [5.8] vs [4.5] -- [8.0] for pre-main sequence stars with circumstellar disk (Class II). The criteria proposed by them also filters out extra-galactic contaminants from the source list. We identify 5 Class I and 20 Class II candidate YSOs. 
\end{enumerate}

\begin{figure}
\centering
\includegraphics[scale=0.32,bb=70 10 900 720,clip]{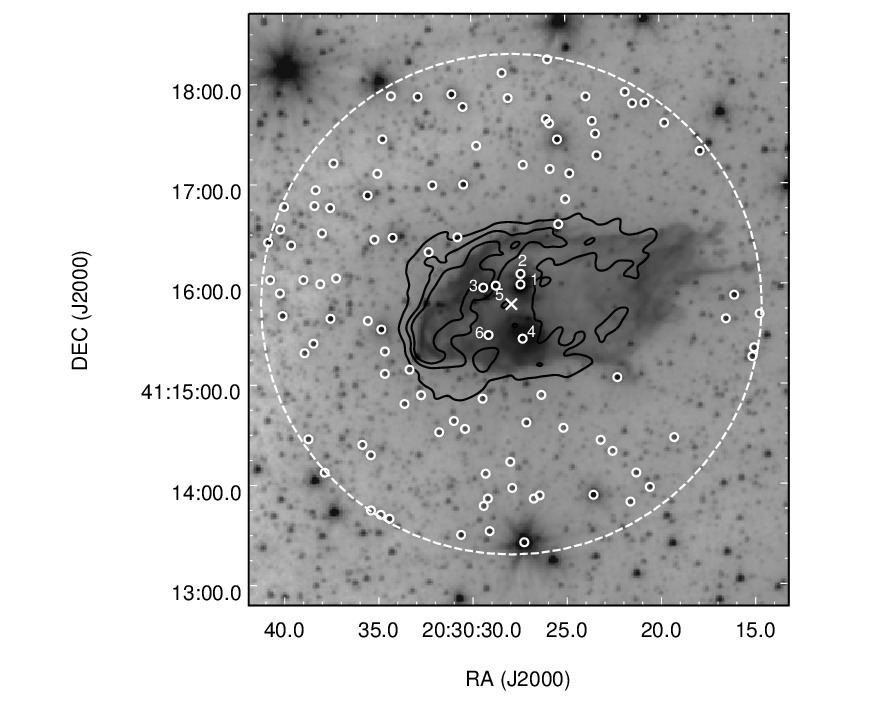}
\caption{The distribution of YSOs (white open circles) are marked on {\it Spitzer} 3.6 $\rm \mu m$ band image. 
Overlaid on the image are the 1280 MHz contours (generated using the convolved low
resolution map, see Section \ref{ionized}). The contour levels are at 5, 10 and 15 times the
$\sigma$ (0.187 mJy/beam) level. The large dashed white circle shows the 150\arcsec radius
region, centered on the IRAS point source position, used for studying the YSO population.
The $\times$  mark denotes the position of the IRAS point source.}
\label{allyso}
\end{figure}

The CC diagrams for the various identification schemes are shown in Figure \ref{spz_class}.
Using the IRAC colors and based on the above four schemes, we identify a total of 78 YSOs 
within 150$\arcsec$ of IRAS 20286+4105. Figure 
\ref{allyso} shows the distribution of the identified YSOs, which are highlighted on the 3.6~$\rm \mu m$ IRAC
image. The figure also shows the 1280~MHz radio contours (see Section \ref{ionized}). As seen in the figure, YSOs are predominantly 
distributed towards the east beyond the arc-like ionized emission. A possible picture of
triggered star formation is explored in Section \ref{ionized}. 
All the YSOs located inside the IRAS 20286+4105 cloud are found to be
Class I sources which appear to cluster near the center of the cloud close
to the position of the IRAS point source. These sources are classified as Class I YSOs in atleast one of the four 
schemes. Other parts of the cloud do not seem to harbour any YSOs. The Class I YSOs located inside the cloud are 
numbered from 1 to 6 and listed in Table \ref{tab_sed}. Sources 1, 2, 3, and 4 are the sources A, B, C, and D of \citet{2010MNRAS.404..661V}. 
These sources are discussed by them to be surrounded by nebulosities and showing 
reddening and/or IR excess. It should be noted here
that sources 1 and 4 do not have available IRAC catalog magnitudes in the 5.8 and 8.0~$\rm \mu m$ bands. 
The magnitudes used here are estimated using the IRAF task `qphot'. 

The MIPS 24~$\rm \mu m$ image shows the presence of two nearly spherical, bright regions with 
the northern one (which is saturated in the core) being extended to the east (see Section \ref{dust} and figure therein). 
Sources 1 and 2 are associated with the northern region and source 4 is associated with the southern region. The eastern extension of the northern region includes 
sources 3 and 5. No 24~$\rm \mu m$ emission is seen towards source 6. 
  
\subsubsection{Spectral energy distribution of selected YSOs}
\label{sed}

In order to determine the stellar parameters of the six YSOs (Sources 1 to 6) located inside the cloud, we 
model the spectral energy distributions (SEDs) using the command-line version of the SED fitting tool of \citet{2007ApJS..169..328R} which
uses YSO models from \citet{2006ApJS..167..256R}. The YSO models are computed by
adopting Monte Carlo based radiative transfer algorithms which use various combinations of central star, disk, infalling envelope, cavities carved out by bipolar outflows. A reasonably large parameter space is explored in these models. 

We use NIR, MIR and FIR data from 2MASS\footnote{This publication makes use of data products
from the Two Micron All Sky Survey, which is a joint project of the University of Massachusetts and the Infrared
Processing and Analysis Center/California Institute of Technology, funded by the NASA and the NSF.}, UKIDSS (for source 5), {\it Spitzer}-IRAC, 
{\it Spitzer}-MIPS, WISE\footnote{This publication makes use of data products from the Wide-field Infrared Survey Explorer, which is a joint project of the University of California, Los Angeles, and the Jet Propulsion Laboratory/California Institute of Technology, funded by the National Aeronautics and Space Administration.}(22~$\rm \mu m$), and {\it Herschel}.
Sources 1, 4, and 6 have WISE point source
counterparts. Source 2 lies at an angular distance of 6\arcsec~from source 1.
We use the WISE 22~$\rm \mu m$ point source magnitudes 
as upper limits given the poor resolution of the images.  From the radial profiles of
isolated point sources in the 22~$\rm \mu m$ image, we estimate the FWHM to be $\sim 18\arcsec$. For sources 1 and 2 we use the same upper 
limit as the separation between them is less than the resolution of the image.
{\it Spitzer}-MIPS 24~$\rm \mu m$  point source magnitude for source 3 is also taken as an upper limit given the contamination 
from source 5. For source
5, the 24~$\rm \mu m$ upper limit is estimated by extracting the flux within an appropriate aperture centered on the source. For all the sources, 
flux densities are also estimated from the {\it Herschel} maps and used as upper limits in the SED fitting.  
We have included a conservative 10\% error on the fluxes while fitting the SED models.
Distance and visual extinction
are taken as free parameters in the models. We adopt a distance range from 1.4 - 1.8~kpc. We 
estimate the extinction by de-reddening the stars using the JHK colour-colour plot \citep{{2003ApJ...598.1107K},{2004ApJ...608..797O}}. 
This is done by shifting them along the reddening vector 
to a line drawn tangential to the turn-off point of the main sequence locus \citep{2006A&A...452..203T}.
The estimated values range from A$ _{v} = 0 $ to 20 mag and this range is used for the model fitting.
   
\begin{figure*}
\centering
\includegraphics[scale=0.4]{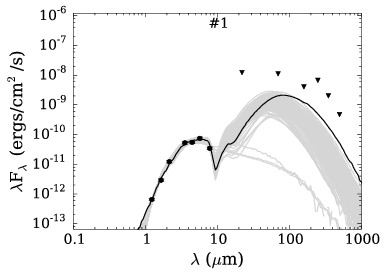}
\includegraphics[scale=0.4]{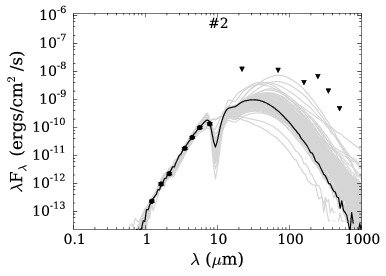}
\includegraphics[scale=0.4]{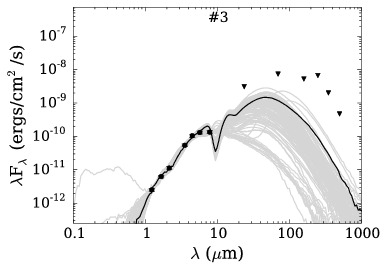}
\includegraphics[scale=0.4]{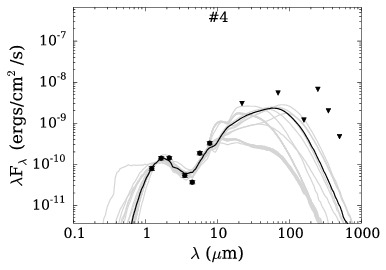}
\includegraphics[scale=0.4]{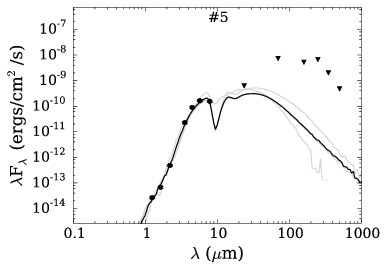}
\includegraphics[scale=0.4]{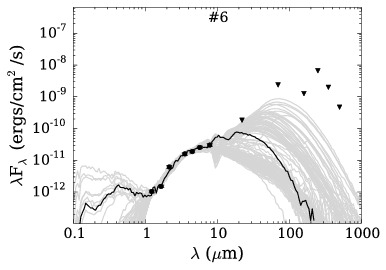}
\caption{SED fitting of the six Class I YSOs using the command-line version of the SED fitting tool of \citet{2007ApJS..169..328R}. The filled circles show the input fluxes and filled triangles are the upper limits, black line shows the best fit and gray lines show the subsequent good fits for $  \chi^{2} - \chi^{2}_{\rm best} < 3$.}
\label{robi}
\end{figure*}

The SED fitting tool gives the best fit model as well as a set of well fit models ranked by 
their $ \chi^{2} $ values. We considered only those models which satisfy
\begin{center}
$  \chi^{2} - \chi^{2}_{\rm best} < 3$ (per data point)
\end{center}
The resulting SEDs are shown in Figure \ref{robi}. The weighted (inverse $\chi^{2} $ of each 
model is taken as the weight) mean and standard
deviations of the parameters obtained for these models are listed in Table \ref{tab_sed}. 
From the model fitted parameters, majority of the Class I YSOs located inside the cloud are found to be intermediate mass stars with 
the derived masses ranging between $\rm 2.5~M_{\odot} $ to $\rm 5.0~M_{\odot} $. Source 5 fits to a
low-mass star of mass $\rm 1.7~M_{\odot} $. Given the models
used and the large parameter space explored with few data points, the derived values of the various parameters should be treated as indicative only.

\begin{table*}
\tiny
\centering
\caption{ Stellar parameters obtained for the six sources located inside the cloud. The values
are the weighted mean and standard deviations of the model output parameters obtained using 
the command-line version of the SED fitting tool of \citet{2007ApJS..169..328R}. The values retrieved from the best fitting model
is given in parenthesis.}
\begin{tabular}{cccccccccc}
\hline
\\
$\#$ & RA & DEC  & log t$ _{*}$ & M$ _{*} $ & log M$_{\rm disk}$ & log $\dot{M} _{\rm disk}$ & log T$ _{*} $ & log L$ _{\rm Total}$& A$ _{v} $\\

&(J2000)  &(J2000)  & (yr)& (M$ _{\odot} $)& (M$ _{\odot} $)& (M$ _{\odot} $/yr)&  (K) & $({\rm L}_{\odot}) $& (mag)
\\
\hline
\\

1 & 20:30:27.4 & 41:15:59.8 & 4.92 $ \pm $ 0.77  &  4.69 $ \pm $ 1.52  & -1.64 $ \pm $ 0.79   & -6.92 $ \pm $ 1.13  & 3.71 $ \pm $ 0.14   & 2.21 $ \pm $ 0.27    & 11.35 $ \pm $ 4.96   \\                                          
& & & (4.63) & (6.02) & (-2.01) & (-8.32) & (3.65) & (2.33) & (6.27) \\

2 & 20:30:27.4 & 41:16:06.0 & 3.84 $ \pm $ 0.72   & 3.06 $ \pm $ 1.83   & -1.64 $ \pm $ 0.74   &  -5.84 $ \pm $ 1.19 & 3.64 $ \pm $ 0.13 & 2.16 $ \pm $ 0.33  &  10.01 $ \pm $ 4.11    \\
& & & (3.49) & (2.80) & (-2.31) & (-6.35) & (3.62) & (2.14) & (4.12) \\
                                                           
3 & 20:30:29.4 & 41:15:57.6 & 4.61 $ \pm $ 1.32 & 3.56 $ \pm $ 1.68 & -1.93 $ \pm $ 1.06 & -6.72 $ \pm $ 1.59 &  3.77 $ \pm $ 0.26 & 2.19 $ \pm $ 0.34 &  7.73 $ \pm $ 4.91  \\
& & & (3.65) & (2.74) & (-1.53) & (-6.21) & (3.62) & (2.11) & (0.85) \\

4 & 20:30:27.3  & 41:15:27.3 & 5.20 $ \pm $ 0.60 &  4.98 $ \pm $ 0.97 & -1.98 $ \pm $ 1.03 & -6.61 $ \pm $ 1.03 & 3.77 $ \pm $ 0.10 & 2.36 $ \pm $ 0.19 & 6.52 $ \pm $ 3.17    \\
& & & (4.66) & (3.73) & (-1.38) & (-5.03) & (3.64) & (2.19) & (6.22) \\

5 & 20:30:28.7 & 41:15:59.1 & 3.89 $ \pm $ 1.41 & 1.69 $ \pm $ 2.28 &  -2.02 $ \pm $ 0.39 &  -5.65 $ \pm $ 1.78 & 3.70 $ \pm $ 0.30 & 1.87 $ \pm $ 0.58 & 11.56  $ \pm $ 3.27  \\
& & & (3.42) & (0.44) & (-1.74) & (-4.83) & (3.56) & (1.53) & (9.67)  \\

6& 20:30:29.1 & 41:15:29.6 & 5.66 $ \pm $ 1.31 & 2.49  $ \pm $ 1.29 &  -2.45 $ \pm $ 0.90 &  -7.66 $ \pm $ 1.38 & 3.88 $ \pm $ 0.23 & 1.69 $ \pm $ 0.41 &  9.04  $ \pm $ 6.34   \\
& & & (6.77) & (4.67) & (-2.85) & (-8.9) & (4.18) & (2.49) & (1.63) \\
\\
\hline         
\end{tabular}
\label{tab_sed}
\end{table*}

\subsection{Emission from ionized gas}
\label{ionized}
\begin{figure*}
\begin{center}
\includegraphics[height=5.5cm,width=8.5cm]{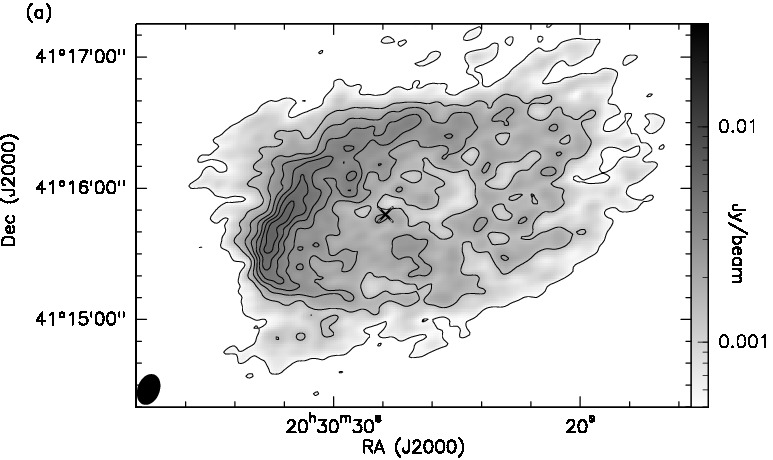} 
\includegraphics[height=5.5cm,width=8.5cm]{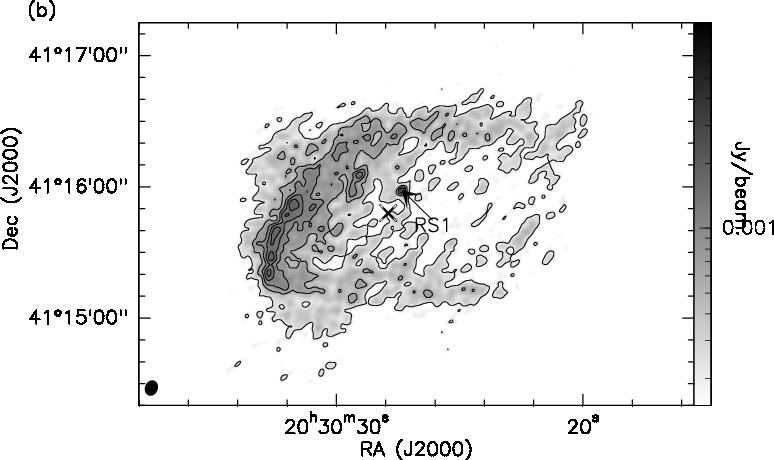} 
\includegraphics[height=5.5cm,width=8.5cm]{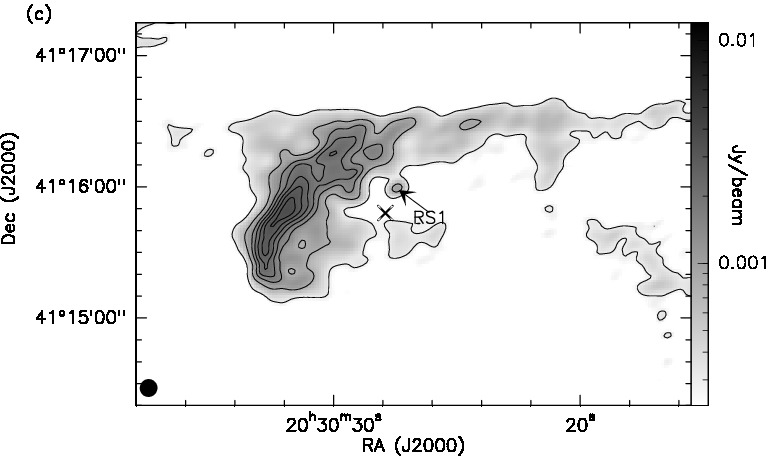} 
\includegraphics[height=5.5cm,width=8.5cm]{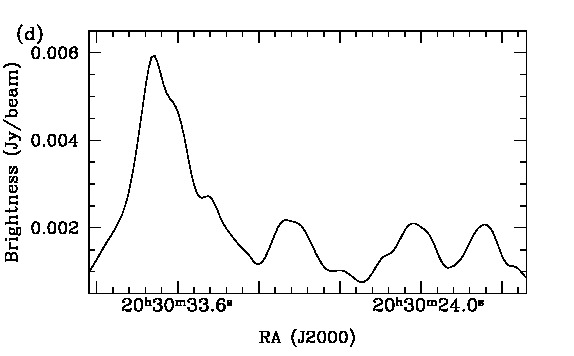} 
\end{center}
\caption{Radio continuum maps at (a) 610 MHz (b) 1280 MHz and (c) 4.89 GHz for the region around IRAS 20286+4105. The contour levels are  at 3, 8, 13, and 18 $\sigma$ levels, where $ \sigma $ is the rms noise in the maps. For
the 610~MHz map 23, 28, and 33 $\sigma$ levels are also shown. The IRAS source position is marked with a $\times$ symbol. The intensity profile along 90$^{\circ}$ PA (taken east of north) at $\delta$ = $\rm 41^{\circ}15\arcmin 48\arcsec$ of radio emission at 610~MHz is shown in (d).}
\label{radiomap}
\end{figure*}

The radio continuum maps probing the ionized gas associated with IRAS 20286+4105 at 610~MHz, 1280~MHz and 4.89~GHz are shown in Figure \ref{radiomap} and the details of the maps are given in Table \ref{radiotab}. 
\begin{table*}
\centering
\caption{Details of the radio continuum maps from GMRT and VLA. The integrated flux values are upto the 3$ \sigma $ level, where $ \sigma $ is the rms noise in the maps.}
\begin{tabular}{|l|c|c|c|}
\hline
Details & 610 MHz & 1280 MHz & 4.89 GHz \\
\hline
Date of Observation & 09 April 2004 & 28 September 2002 & 15 April 1988\\
 
Synth. beam &  7$\arcsec$.26 $  \times$   5\arcsec.04 & 2\arcsec.91 $ \times $   2\arcsec.26 & 4\arcsec.07 $ \times $   4\arcsec.01\\
 
Integrated Flux (mJy)&  781 & 508 & 150\\

rms noise (mJy/beam) &  0.170 & 0.070 & 0.100\\
\hline
\label{radiotab}
\end{tabular}
\end{table*}
The radio maps clearly show a bright arc-like emission toward the east and extending northwards with a 
decreasing intensity distribution in the south-west direction. 
The  GMRT maps show the presence of diffuse emission in the entire cloud. The diffuse emission is more extended at 610~MHz compared to the
1280~MHz map. The 4.89~GHz VLA map
mostly shows the bright arc. The intensity variation of the ionized region along a central declination 
is also shown in Figure \ref{radiomap} (d). The profile resembles the characteristic 
intensity distribution of a cometary HII region as described in \citet{1989ApJS...69..831W}. 
We discuss the cometary nature of the radio morphology in detail toward the end of this section.
 
Apart from the diffuse emission, the 1280~MHz map peak coincides with a point-like compact knot
(hereafter RS1) located at $\alpha_{2000}$=$\rm 20^h30^m27.34^s$; $\delta_{2000}$=$+41^{\circ}$15\arcmin58.51\arcsec. This knot 
is also clearly seen in the 4.89~GHz map.  
The position of this knot is toward the centre of the cloud and west of the bright, diffuse arc-type emission. 
Indication of this radio knot can also be seen in the map presented by \citet{1991AJ....101.1435M} (see their
Fig. 1). However in the 610~MHz map, RS1 is not as prominent given the dominant diffuse emission at this frequency. 
A more detailed study of RS1 is presented in the next section.

The physical properties of the ionized gas in this region are derived from the flux density measurements 
obtained from the radio maps.
The number of hydrogen ionizing photons (h$ \nu $ $ > $ 13.6 eV) per second required to maintain 
the ionization of the nebula and hence the spectral type of the ionizing star is estimated using the 
1280~MHz flux density and the formulations discussed in \citet{2003A&A...407..957M} and 
\citet{1973AJ.....78..929P}. Assuming the emission to be optically thin at this frequency and 
a single ionizing source responsible for the same, the number of Lyman continuum photons is estimated using the 
following expression from \citet{2003A&A...407..957M},
\begin{center}
\begin{equation}
\begin{aligned}
 N_{\rm lyc} = 4.634 \times 10^{46} \left(\dfrac{T_{e}}{10^{4} {\rm K}}\right)^{-0.45} \left(\dfrac{\nu}{{\rm 5~GHz}}\right)^{0.1} \left(\dfrac{D}{{\rm 1~kpc}}\right)^{2}  \\
 \times \left(\dfrac{S_{\nu}}{{\rm 1~Jy}}\right) a_{\nu}^{-1} [{\rm s^{-1}}]
\end{aligned}
\end{equation}
\end{center}
where, $T_{e} $ is the electron temperature, $ \nu $ is the frequency in GHz, $S_{\nu} $ is the
integrated flux, $D$ is the heliocentric distance and $a_{\nu} $ is the correction factor (taken as unity similar to \citet{2003A&A...407..957M}). $T_{e} $ is 
estimated to be $ \sim $ 7300 K using the following equation from \citet{2000MNRAS.311..329D}
\begin{equation}
T_{e}[K] = (372 \pm 38) D_{G} + 4260 \pm 350 
\end{equation}
where $D _{G} $ is the Galactocentric distance of the source which is determined to be 8.16 kpc following
the expression from \citet{xue2008milky}. 
We estimate log$ _{10} $($N_{\rm lyc} $) to be 46.79 which translates to a main-sequence spectral type between B0.5 -- B0 (see Table 2 of \citet{1973AJ.....78..929P}). 
The spectral type determined from this work is within one subclass of the estimates obtained 
from previous radio continuum studies by \citet{1989ApJ...345L..47O} and 
\citet{1991AJ....101.1435M} after scaling to the distance and electron temperature used in our 
study. \citet{2003A&A...398..589V} carried out the radiative transfer modeling
of two FIR cores located towards the cloud centre (see Section \ref{dust}) using MIR and FIR 
data and estimated the spectral type of the sources associated with these 
cores to be between B0.5 - B0. Their modeling predicted no radio emission. 
It should be noted here that the GMRT maps do not show any radio peaks associated with the 
cores except the radio knot RS1 even though diffuse radio emission is detected in the entire 
cloud.

We estimate the emission measure and the electron density of the ionized region associated
with IRAS 20286+4105 by assuming the radio emission to be free-free from an isothermal, spherically symmetric and homogeneous medium \citep{{2006A&A...452..203T},{2014MNRAS.440.3078V}}. For this, 
we convolve the 4.89~GHz and 1280~MHz maps to the 
resolution of 610~MHz map which has the lowest resolution ($7\arcsec.26 \times 5\arcsec.04$). 
The extent of the eastern, bright radio arc is found to be 
roughly 15 - 20\arcsec~wide if we include only the bright emission region (above 15$\sigma$). The
peaks of these convolved radio maps are located within this bright arc. The 1280~MHz map shows two 
discrete peaks (5.76 mJy/beam at $\alpha_{2000}$=$\rm 20^h30^m31.69^s$; $\delta_{2000}$=$\rm +41^{\circ}$15\arcmin51.51\arcsec and 5.47 mJy/beam at 
$\alpha_{2000}$=$20^h30^m32.43^s$; $\delta$=$\rm +41^{\circ}$15\arcmin38.20\arcsec) on the arc. The 4.89~GHz map shows a single peak which 
matches within $\sim 3\arcsec$ of the brighter 1280~MHz peak. The 610~MHz map does not clearly display any
discrete peak and shows a rather elongated flux enhanced region. The location of the peak flux density lies within 
$\sim 1.5\arcsec$ of the secondary peak of 1280~MHz. From the 1280~MHz and 4.89~GHz peak fluxes, we derive 
the emission measure and the electron density to be $\rm 1.75 \times 10^5 
\,cm^{-6}\, pc$ and $\rm 1.9 \times 10^3\, cm^{-3}$, respectively, for the estimated electron temperature of 7300~K.

\begin{figure}
\centering
\includegraphics[scale=0.3]{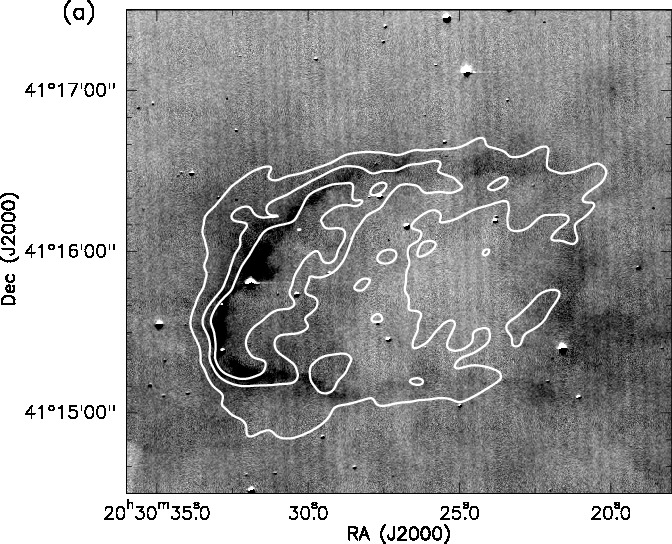}
\includegraphics[scale=0.3]{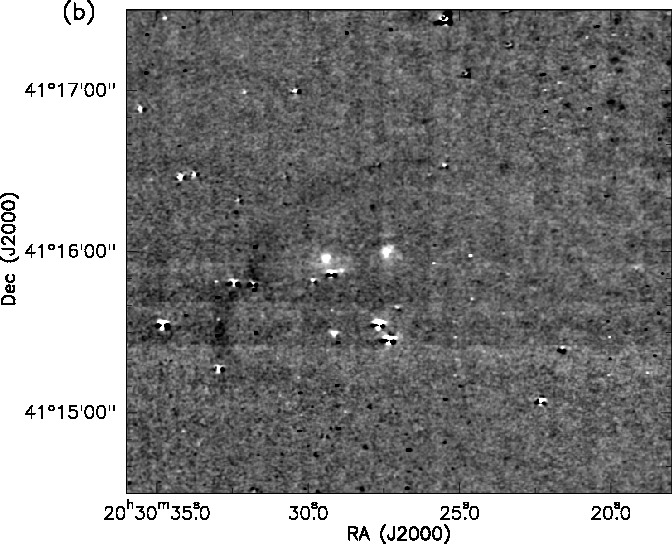} 
\caption{Ionized emission in the (a) H$_{\alpha}$ and (b) Br$\gamma$ for the region associated with IRAS 20286+4105. Overlaid on the H$_{\alpha}$ image are the 1280~MHz contours (generated using the convolved low resolution map). The
contour levels are at 5, 10 and 15 times the $\sigma$ (0.187~mJy/beam) level.
Residuals of continuum subtraction are seen in the images.}
\label{brha}
\end{figure}

The ionized emission is also probed in the optical and NIR using H$\alpha$ (656.8~nm) and 
Br$\gamma$ (2.166~$\rm \mu m$) emission lines. 
Figure \ref{brha} shows the continuum-subtracted H$\alpha$ and Br$\gamma$ images.
The signal-to-noise ratio in the continuum-subtracted Br$\gamma$ image is rather
poor. Nevertheless, we clearly detect a faint arc of emission. The H$\alpha$ image 
shows a much more extended arc, whereas the Br$\gamma$ emission is prominent mostly towards the southern
part of the arc. However, in the absence of flux calibrated images, it is difficult to decouple the effect of 
instrument sensitivities. We have overlaid 1280~MHz radio contours (generated using 
the convolved low resolution map) on the H$\alpha$ line emission image.  This shows the correlation between the
ionized components at optical, infrared and radio wavelengths.
The optical and NIR continuum-subtracted line images trace the arc-like radio morphology. 
However, since the radio emission is least affected by extinction, the maps  
detect the extended, faint, and diffuse emission spread over the entire cloud associated with IRAS 20286+4105. 
 
A 2MASS source (J20303183+4114548: J = 11.845 mag; H = 11.587 mag;
K$_s$ = 11.458 mag) with an optical counterpart (SDSS r magnitude = 14.445 ) lies within 4\arcsec~of the brighter
1280~MHz peak. 
The second radio peak is also seen to be within 4\arcsec~of another 2MASS source (J20303274+411540 - J = 15.795; 
H = 14.849; K$_s$ = 14.568) with an optical counterpart (SDSS r magnitude = 19.9 mag). 
The NIR magnitudes and colours of both these sources, which lie within the radio arc, suggest that these are likely to be main-sequence field sources in the line of sight. 
Apart from these, the nearest identified IRAC YSO is located $\sim 9\arcsec$ towards the east of 
the radio peaks and hence unlikely to be associated with them.  

The observed radio arc points towards
the direction of the Cygnus OB2 cluster. IRAS 20286+4105
cloud falls within the radius of influence of this cluster \citep{{2006A&A...458..855S},{2011ApJ...727..114R}} being located at a distance of $\sim$20~pc from the
cluster centre. Based on the age of the Cyg OB2 cluster
and the distance to it, it is possible that the ionization front emanating from the 
massive stellar population of the cluster would have reached and ionized the IRAS 20286+4105 cloud. From the mass, luminosity and the 
geometry of the IRAS 20286+4105 cloud, \citet{2011ApJ...727..114R} suggest that the radio emission 
is possibly due to this external ionization field. It should be noted here that the authors do not give any quantitative proof for the above inference.

\begin{figure}
\centering
\includegraphics[scale=0.32,bb=35 10 900 530,clip]{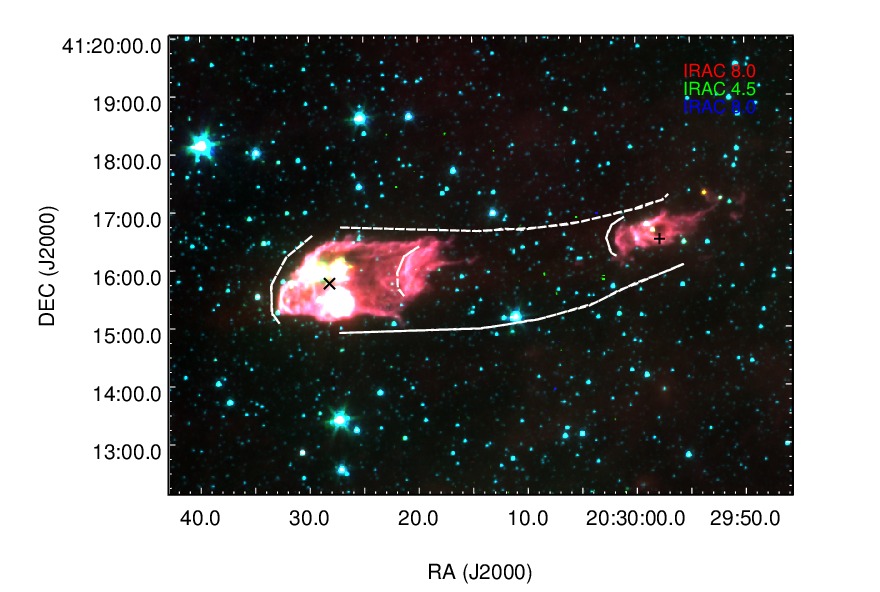}
\caption{{\it Spitzer} IRAC 3 - color composite image of the region around IRAS 20286+4105. The position of IRAS source and 
SNR G079.8+01.2 
are marked with 'x' and '+', respectively. The figure also marks the curvatures and the northern and southern edges based 
on visual inspection.}
\label{snr}
\end{figure}

However, given the cometary morphology of the HII region associated with IRAS 20286+4105 
and the presence of a nearby supernova remnant, SNR G079.8+01.2 (IRAS 20281+4106) 
\citep{2003ApJS..149..123S}, we attempt to explore a different angle where a likely association
of both is invoked.  In Figure \ref{snr}, we show the relative location of SNR G079.8+01.2 and the
IRAS 20286+4105 cloud in the plane of the sky. SNR G079.8+01.2 lies in the region belonging 
to the Cygnus OB2 association at an angular distance of $\sim 6.5\arcmin$ from the position of the radio 
peak of IRAS 20286+4105 cloud. Based on the IRAS flux densities, \citet{1992A&A...266..202P} have
identified IRAS 20281+4106, associated with the SNR, as a possible Class I YSO. 
These authors have assumed IRAS 20281+4106 to belong to the 
Cygnus OB2 association ($\rm d = 1.7~kpc$). 
Using the $\rm HCO^{+}$ line velocity measurements given in \citet{2013ApJS..209....2S}
and the rotation curve of \citet{1993A&A...275...67B},
we estimate the kinematic distances to IRAS 20286+4105 and SNR G079.8+01.2 to be 3.8 and 4.1~kpc, respectively. Based on these estimates, we assume the SNR to be at the same distance as the
IRAS 20286+4105 cloud. However, we adopt the distance of 1.61~kpc for both the complexes.

Focusing on the morphology of the HII region, there are several models proposed in 
literature for the mechanisms behind such morphology that is commonly observed in HII regions. 
\citet{1985ApJ...288L..17R} were the first to propose the relative motion between the ionizing 
source and the extended molecular environment. This led to the `bow-shock' model 
\citep{{1990ApJ...353..570V}, {1991ApJ...369..395M}, {1992ApJ...394..534V}}, which requires highly 
supersonic motion of a wind-blowing, ionizing star through dense surrounding material. 
The other frequently adopted model is the `champagne-flow' model which assumes the ionizing star 
to be nearly stationary with respect to the molecular cloud. The cometary morphology
in this case is attributed to the steep density gradient existing in the surrounding medium 
\citep{{1978A&A....70..769I}, {1979A&A....71...59T}}. The column density map derived using {\it 
Herschel} data reveal the presence of a high-density clump in the middle of the IRAS 20286+4105 
cloud (see Section \ref{dust}). 
The ionized gas distribution shows a decrease in intensity in the south-west direction towards the 
clump which is opposite to what is expected from the `champagne-flow' model and hence
this model is unlikely to explain the observed cometary morphology.

The occurrence of a bow-shock in the IRAS 20286+4105 cloud requires a star with large
velocity generally termed as a `runaway' star.   
Two mechanisms are generally proposed to explain the origin of their high velocities. 
One of these is the binary supernova scenario where the `runaway' star is initially part of a binary system. It acquires high velocities when 
its binary companion explodes as a supernova \citep{{1961BAN....15..265B},{1998A&A...331..949M}}.
The other mechanism involves dynamical ejection from a dense cluster \citep{{1967BOTT....4...86P},
{1986ApJS...61..419G}}. \citet{2001A&A...365...49H} give evidences for both mechanisms based
on accurate astrometry observations. Our aim is to first quantitatively attribute the bow-shock
mechanism with the observed cometary morphology. Subsequently, we try to probe the likely 
connection between the SNR and the HII region.

A `bow-shock' is detected at the stand-off distance ($\rm r_s$) from the ionizing 
star where the ram pressure and momentum flux of the wind and the interstellar medium (ISM) 
balance 
 \citep{{1990ApJ...353..570V}, {1991ApJ...369..395M}}. Equating these quantities, the
stand-off distance is shown to be \citep{1988ApJ...329L..93V},
\begin{equation}
\label{r_s}
r_{s}=1.78\times 10^{3}\sqrt{\frac{\dot{M}v_{\infty }}{\mu _{\rm H}n_{\rm H}v_{\rm \star-ISM}^{2}}}\ {\rm pc}
\end{equation}
where, $\dot{M}$ is mass-loss rate from star in M$_{\odot}\,\rm yr^{-1}$ and $v_{\infty}$ is the 
terminal velocity of the stellar wind in $\rm  km~s^{-1}$. $\mu _{\rm H}$ and $n_{\rm H} $ are the
mean mass per hydrogen nucleus and the hydrogen gas density in $\rm cm^{-3}$, respectively. $v_{\star-\rm ISM}$ is the velocity 
of star with respect to ISM in km $\rm s^{-1}$. The mass-loss rate and the terminal velocity are obtained using the 
following equations from \citet{1991ApJ...369..395M}.
\begin{equation}
\label{m0}
\dot{m}=2.0\times 10^{-7}\left ({L}/{L_{\odot }}\right )^{1.25}  \\
\end{equation}
\begin{equation}
\label{vw}
{\rm log}\ v^{\prime}_{\infty}=-38.2+16.23\ {\rm log} \ T_{{\rm eff}}-1.70\ ({\rm log}\ T_{{\rm eff}})^{2}
\end{equation}
where, $L$ is the stellar luminosity, $L_{\odot }$ is the solar luminosity and 
$T_{{\rm eff}}$ is the effective temperature of star. In Equation \ref{r_s}, $\dot{M} = \dot{m} \times 10^{-6}\,{\rm M}_\odot$ 
$\rm yr^{-1}$, $v_{\infty } = v^{\prime}_{\infty} \times 10^3 \,\rm km\, s^{-1}$. 
The luminosity and effective temperature are $2 \times 10^4 - 4.8 \times 10^4 L_{\odot}$                                                                                                                                                                                                                                                                                                                                                                                                                                                                                                                                                                                                                                                                                                                                                                                                                                                                                                                                                                                                                                                                                                                                                                                                                                                                                                                                                                                                                                                                                                                                                                                                                                                                                                                                                                                                                                                                                                 
and 26200 - 30900 K, respectively for the estimated spectral type of B0.5 - B0 \citep{1973AJ.....78..929P}. 
It is to be noted here that the recently released 
parallax catalog from URAT \citep{2016AJ....151..160F} has the astrometry of most 2MASS sources
in the Cygnus region. However, in our case, we do not have any 2MASS counterpart of the radio
peak which is the likely position of the ionizing star. Thus, we lack information regarding 
the astrometry of the ionizing star.  
Hence, for our calculations, we adopt a typical space velocity of 
$\rm \sim 10~km \, s^{-1}$ \citep{1992ApJ...394..534V} for the supersonically moving ionizing 
star. Plugging in these values in the above equations, we determine the mass-loss
rate, $ \dot{M} = 2.2 - 6.3 \times 10^{-8}\,{\rm M}_\odot$ $\rm yr^{-1}$ and the terminal velocity $v_{\infty} = 2.1 - 2.5 \times 
10^3\,{\rm km s^{-1}}$.
 
The medium through which the star moves would be a combination of ionized and neutral medium. 
If we consider the HII region to be fully ionized, then the electron
density, $n_{\rm e} = n_{\rm H}$.
Taking  $\rm \mu _{H} = 0.61$ (for a fully ionized medium) and using
Equation \ref{r_s}, we estimate the stand-off distance to lie between 0.05 - 0.09~pc which translates
to 6.8\arcsec - 12.6\arcsec. However, if we consider a neutral medium, 
then we use $n_{\rm H} = 1.2 \times 10^4\, \rm cm^{-3}$
derived from the column density maps for Clump 1 
(refer to Section \ref{dust}) and $\mu _{\rm H} = 1.4$. In this
case, the stand-off distance lies between 0.01 - 0.03~pc or 1.8\arcsec - 3.4\arcsec.
From the 1280~MHz radio map, the angular separation between the peak and the `edge' of the bright emission is 
estimated to be $\sim$ 7\arcsec~. This is consistent with the stand-off
distance calculated for the fully ionized case thus supporting the bow-shock scenario.
A further point to note is regarding
the H$_{2}$ line emission of the region associated with IRAS 20286+4105 which is
presented in \citet{2010MNRAS.404..661V}. The H$_{2}$ line image also shows the
presence of an arc, the location of which matches with the 
position of the radio arc seen in our maps. Morphologically, the H$_{2}$ line emission 
is seen to be irregular compared to the radio emission. Considering the above bow-shock
scenario, it is likely that this H$_{2}$ line emission is shock-excited . 

The morphology of the marked arcs and the 
shape of the northern and southern edges of the IRAS 20286+4105 cloud and the SNR, as highlighted
in Figure \ref{snr}, indicates their likely association. The curvature of the arcs points in the 
direction from the SNR to the cloud.
Going by the binary supernova model, we expect the SNR to be the parent location of the
star responsible for the bow-shock.
For the assumed velocity
of $\rm 10\,km\,s^{-1}$, we estimate the kinematic age (which is defined to be the time since this runaway star left its
parent association) to be $\rm \sim 3\times10^5 yr$. The kinematic age thus derived can be considered as an upper limit since the initial velocity of the star could be higher compared
to the velocity considered inside the cloud. The derived age is considerably shorter than the main-sequence lifetime
of $\rm (12 - 16.4)\times10^6 yr$ \citep{2011ApJ...730L..33M} of the ionizing star which is estimated to be of spectral type between B0 - B05. 
This difference in the ages is consistent with the 
fact that runaway stars acquire high velocities after an initial evolutionary phase as member of a close binary system in 
which the primary evolves for several million years before it explodes as a supernova and the runaway is ejected out of the
system \citep{2001A&A...365...49H}. 
The above analysis suggests that the radio emission seen in the IRAS 20286+4105 cloud
is due to a possible runaway star which was likely associated with SNR G079.8+01.2
in the past. 

The above picture has implication on the star formation activity in the IRAS 20286+4105 cloud as well. The SED model fitted ages of the six Class I YSOs located inside the cloud range between 
$\rm 7\times10^3 yr$ to $\rm 5\times10^5 yr$. If these are taken as representative values, then
it is possible that this population is the result of a shock-triggered collapse of a pre-existing 
clump by an encroaching shock either from the strong stellar wind of the precursor of the supernova or originating from the energetics of the explosion itself. The collapse could 
also have been induced by the passing runaway star. Further evidence of triggered star formation
is also seen towards the east beyond the radio arc where there is a over density of YSOs. The
above interpretation finds strength in the morphology of IRAS 20286+4105 cloud and the SNR 
discussed earlier.  However, it should be noted here that the 
simplistic calculations and the interpretations given above 
are based on model fitted age parameters and on the visual inspection of the projected morphology
of the complex. With the given data, it is difficult to make further quantitative interpretation.

As is clear from previous discussions, two scenarios emerge related to the ionized
emission associated with the IRAS 20286+4105 cloud. Without disqualifying the possibility of
external ionization due to the massive stars of Cygnus OB2 cluster, we are inclined towards the bow-shock, `runaway' star and the                                                                                                                                                                                                                                                                                                                                                                                                                                                                                                                                                                                                                                                                                                                                                                                                                                                                                                                                                                                                                                                                                                                                                                                                                                                                                                                                                                                                                                                                                                                                                                                                                                                                                                                                                                                                                                                                                                                                                                                                                                                                                                                                                                                                                                                                                                                                                                                                                                   supernova
picture. This is driven by the proximity of SNR G079.8+01.2 and the correlation seen in the 
morphology discussed and shown in Figure \ref{snr}. Accurate distance estimates, identification
of the ionizing star and determination of space velocity are required before we can conclusively
ascertain the mechanism behind ionization in the cloud.
 
\subsection{Nature of radio knot RS1}
\label{rs1}

As mentioned in Section \ref{ionized}, a radio knot, RS1, is visible towards the center of the complex around 
10\arcsec~north of the IRAS point source. RS1 is located at the 
peak position of the high resolution (2\arcsec.91 $ \times $   2\arcsec.26) 1280~MHz map and 
lies within $\sim$ 1.5\arcsec~of Source A and of 7\arcsec~of Source B of \citet{2010MNRAS.404..661V}.
From narrow-band $\rm H_{2}$ observations, \citet{2010MNRAS.404..661V} show the presence of two outflows
associated with this region
(see Fig. A40 of their paper). They suggest a deeply embedded source north of Source A to be responsible for
the east-west aligned outflow and Source B to be the driving YSO for the other outflow.
Based on the location, it is likely that RS1 is associated with the aligned outflow. Figure \ref{rs1}
shows the position of RS1 in comparison to the outflow directions given in \citet{2010MNRAS.404..661V}. As shown in the 
figure, there are a number of masers detected towards IRAS 20286+4105. The positions of H$ _{2} $O \& NH$ _{3} $ masers 
detected by \citet{2011MNRAS.418.1689U} are coincident with that of RS1. Water masers are known 
for their closest association with outflows from young stars \citep{2005IAUS..227..180D}. 
It should be noted here that in some papers
the IRAS point source position is listed against the detected masers which may not be the actual position of the masers. 
\begin{figure}
\centering
\includegraphics[scale=0.32]{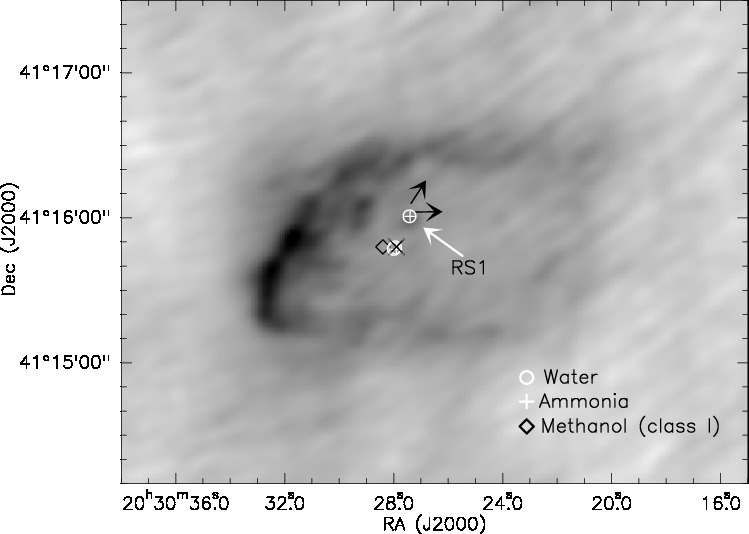}
\caption{Position of RS1 and maser emission spots in the region are marked on the low resolution 1280~MHz radio image.
Also marked as black arrows are the two possible collimated outflows discussed in \citet{2010MNRAS.404..661V}. The $\times$ 
shows the position of the IRAS point source.}
\label{rs1}
\end{figure}

\begin{figure}
\centering
\includegraphics[scale=0.4]{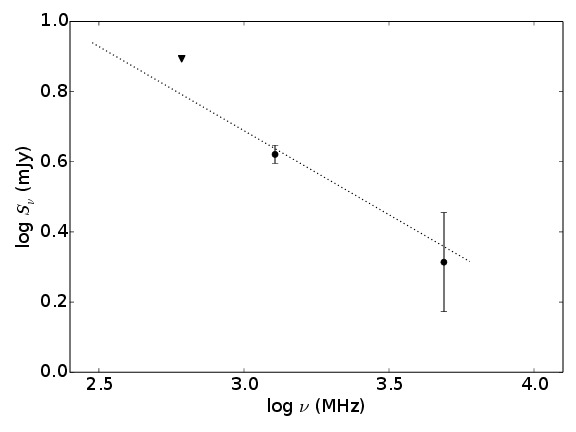} 
\caption{The radio SED of RS1. The dotted line shows the fitted spectral index slope using 
flux densities at 1280~MHz and 4.89~GHz. The upper limit to the flux density at 610 MHz is shown as a 
triangle.}
\label{alphased}
\end{figure}

To probe the radio nature of RS1, we investigate the radio SED obtained using the low resolution ($7\arcsec.26 \times 
5\arcsec.04$) maps. We integrate the flux within a square aperture of size 16\arcsec~covering RS1.
Figure \ref{alphased} shows the SED. The integrated flux obtained for RS1 from 
the 610~MHz map is considered as an upper limit since it includes significant contribution from the 
diffuse emission. The slope of this log-log plot is generally defined as the spectral index denoted 
by $ \alpha $ (assuming $S_{\nu} \propto \nu^{\alpha}$). A spectral index of $\sim -0.5 \pm 0.05$ 
is obtained using the flux densities at 1280~MHz and 4.89~GHz. The fitted slope is shown as a dotted line in 
the figure. The derived negative spectral index is not
consistent with the value of $\sim -0.1$ expected from optically thin free-free emission.
\citet{1993RMxAA..25...23R} show that when only free-free emission and absorption mechanisms are involved, the
spectral index is always  $\geqslant -0.1$ irrespective of
the electron density and temperature distribution. 
Spectral indices below $-0.1$ are 
interpreted as co-existing synchrotron emission components with free-free emission. 
Such negative spectral indices are seen in Herbig-Haro objects which are associated with jets / outflows 
\citep{1993ApJ...416..208M}. 
The above discussion indicates that RS1 is possibly associated with the shocked
non-thermal emission from the 
detected outflow. \citet{1993ASSL..186..311C} explained 
that non-thermal emission associated
with jets and outflows are due to the diffusive shock acceleration at working surfaces of 
jets, which moves into the ambient molecular cloud and boosts up the non-thermal emission over 
thermal free-free emission.  For this mechanism to happen, a high density environment 
with strong magnetic field is required. Non-thermal emission due to stellar winds and associated
jets are also discussed by \citet{1995ApJ...443..238R} who also show the correlation of 
water masers with non-thermal sources.

In the infrared, RS1, which coincides with Source A of \citet{2010MNRAS.404..661V}, is an extreme
red and embedded object with NIR colors of J - K = 6.4 and H - K= 2.63. These colors suggest 
large NIR excess that possibly indicates a  highly extincted protostellar candidate 
(compare with Figure 7 of \citet{2006A&A...452..203T}). The MIR IRAC colors
place it in the location of Class I objects (YSO \# 1) in the CC diagrams, consistent with the JHK colors. The SED 
modelling presented in Section \ref{population} suggests an intermediate-mass star with an estimated mass of $~ \rm 4.7 \,M_{\odot}$ and a disk accretion rate of 1.2$ \times 10^{-7} $ M$ _{\odot} $/yr.

\subsection{Emission from cold dust}
\label{dust}

\begin{figure*}
\centering
\includegraphics[scale=0.55,bb=0 70 792 512,clip]{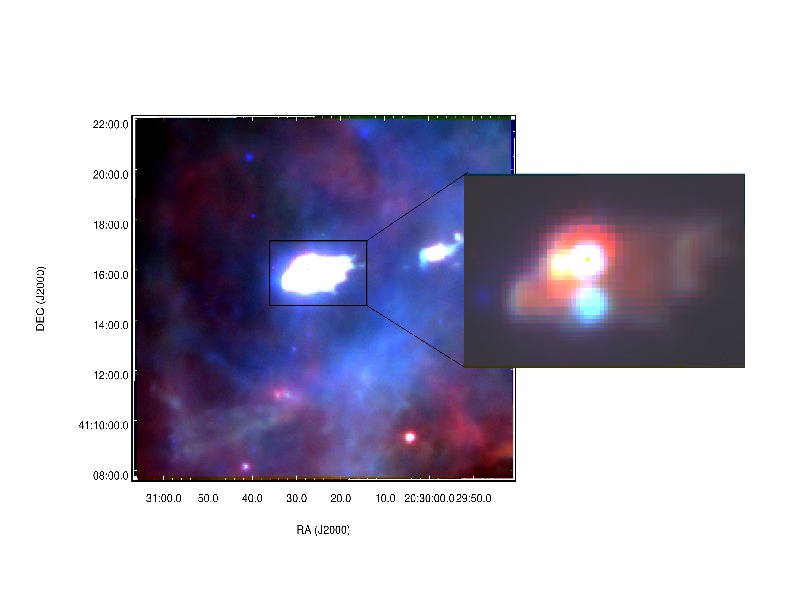} 
\caption{Three color composite image of the region around IRAS 20286+4105 with {\it Herschel}-SPIRE 250~$\rm \mu m$ (red), 
{\it Herschel}-PACS 70~$\rm \mu m $ (green) and 24~$\rm \mu m$ {\it Spitzer}-MIPS (blue) images.}
\label{rgb_herschel_mips}
\end{figure*}

The distribution of dust emission in the region around IRAS 20286+4105 is shown as a colour-composite
image in Figure \ref{rgb_herschel_mips}.
The expanded view shows the complex under study. The image shows the presence of two bright 
nearly spherical regions with the northern one being extended towards
the east. The cold dust environment sampled by 250~$\rm \mu m$ (red) emission is more
extended and dominant towards the northern edge, whereas the warm dust sampled by the 24~$\rm \mu m$ (blue) emission is 
mostly localized in these two 
regions being brighter towards the southern one. These regions are also identified as the northern and 
southern cores and studied by \citet{2003A&A...398..589V}. From the radiative transfer modeling of the MIR and FIR data, 
these authors suggest 
the luminosities of the two cores to be consistent with ZAMS stars of spectral type B0.5 (northern core) and B0 - B0.5 
(southern core). 

The thermal dust emission peaks in the FIR and the Rayleigh-Jeans regime of the cold dust SED
is covered by the {\it Herschel} bands (160 - 500~$\rm \mu m$). In this section we carry out detailed 
study of the physical properties of the cold dust environment associated with IRAS 20286+4105 using
{\it Herschel} data. 

\subsubsection{Temperature and column density distribution}
\label{cdens and temp}

Following the procedure outlined in \citet{2011A&A...535A.128B}, \citet{2012A&A...547A..11N},  \citet{2013A&A...551A..98L}, 
and \citet{2015MNRAS.447.2307M},
a pixel-by-pixel modeling of this dust
emission with a gray body/modified blackbody is adopted in order to generate temperature and column density maps of 
the region of our interest. 
For the preliminary steps, the {\it Herschel} data compatible software HIPE\footnote{The 
software package for {\it Herschel} Interactive Processing Environment (HIPE) is the application that allows users to work 
with the {\it Herschel} data, including finding the data products, interactive analysis, plotting of data, and data 
manipulation.} is used. The image units of PACS ($\rm Jy\,pixel^{-1}$) and SPIRE ($\rm MJy\,Sr^{-1} $) are
different so the first step involves converting the surface brightness unit of the images to $\rm 
Jy\,pixel^{-1}$ using the task `Convert Image Unit'. Further the beam and pixel sizes are different in the five bands and 
for a 
pixel-by-pixel analysis, we need to project the maps onto a common grid with a common pixel size and resolution. 
The plug-in `Photometric Convolution' is used for this purpose. The final images have a resolution of
36$\arcsec$ and pixel size of 14$\arcsec$ which are the parameters of the 500~$\rm \mu m$ image. 

Since the bolometer arrays of {\it Herschel} are not absolute photometers, the maps include sky emission with an unknown 
offset. This sky/background radiation mainly includes cosmic microwave background and the diffuse Galactic 
background. In order to remove the contribution of these background emission components and 
to correct for the offset, background 
subtraction is required. We determine the background flux level from a nearby ($\sim 20\arcmin$ away) 
patch of sky ($\alpha_{2000}$ = 20:31:57.0, $\delta_{2000}$ = +41:27:33), which is relatively free 
of emission in all five bands. The background fluxes in the
bands are estimated by fitting a Gaussian function to the 
distribution of individual pixel values in the selected region. The fitting is done iteratively by rejecting the 
pixel values outside $ \pm 2\sigma $, until the fit converges \citep{{2013A&A...551A..98L}, {2011A&A...535A.128B}, 
{2015MNRAS.447.2307M}}. The resultant 
background flux levels at 70, 160, 250, 350 and 500 $\rm \mu m $ are -1.0, -1.5, 1.1, 0.7 and 0.3 Jy/pixel 
respectively. The negative flux values at 70 and 160 $\rm \mu m $ are due to the arbitrary scaling of the PACS images.

\begin{figure*}
\centering
\includegraphics[scale=0.3]{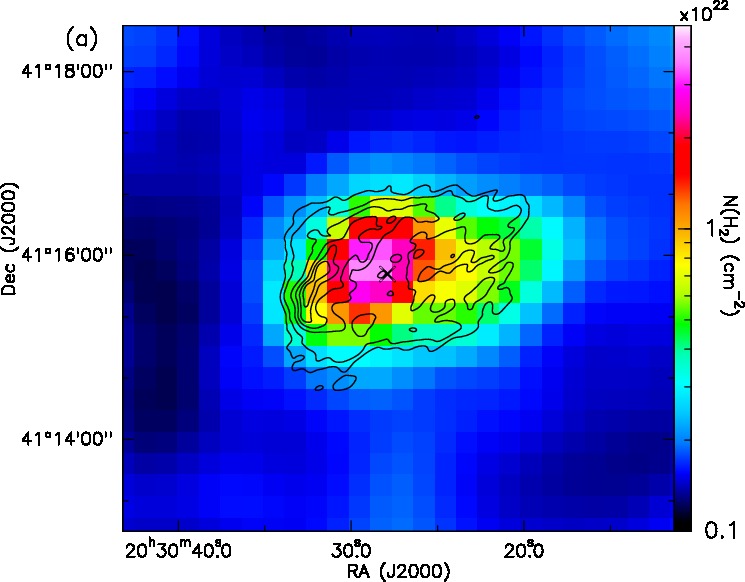}
\includegraphics[scale=0.3]{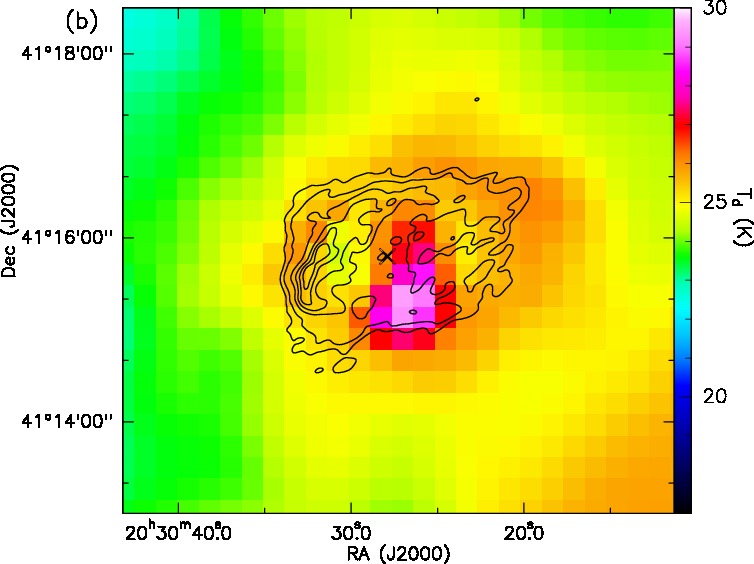}
\includegraphics[scale=0.3]{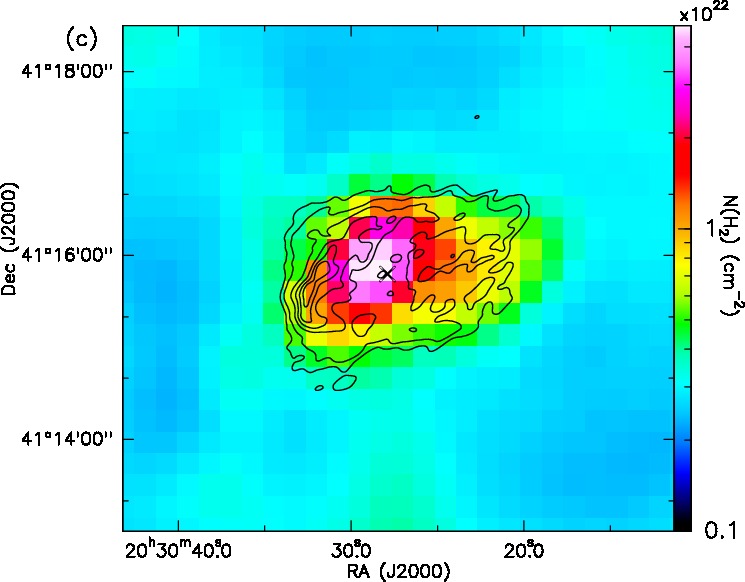}
\includegraphics[scale=0.3]{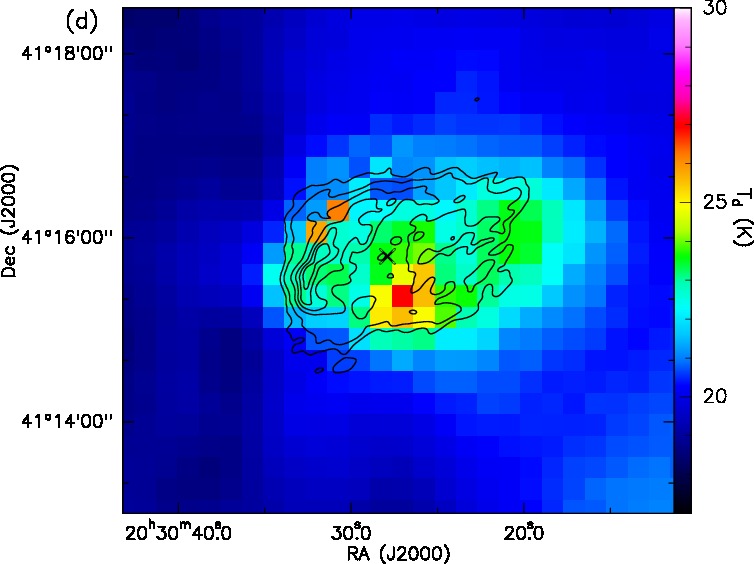}
\caption{(a) Column density and (b) dust temperature map in the region associated with IRAS 20286+4105 derived from 
modified black body fitting using 70 - 500 $\rm \mu m$ data . Overlaid on the images are the low resolution 
1280~MHz radio contours. The contour levels are 3, 5, 10, 15, 20, 25 times $\sigma$ (0.187~mJy/beam).
The $\times$ shows the position of the IRAS point source.
(c) and (d) are same as (a) and (b), respectively but excluding 70~$\rm \mu m$ emission in the modified black body 
modeling.}
\label{temp_cdens}
\end{figure*}

We co-relate the emission to a modified blackbody that takes into consideration the optical depth, and the dust emissivity.  
This is done pixel wise using the following expression  \citep{{1990MNRAS.244..458W}, {2012MNRAS.426..402F}, 
{2013ApJ...766...68P},{2015MNRAS.447.2307M}},
\begin{equation}
S_{\nu}(\nu) - I_{\rm bkg}(\nu) = B_{\nu}(\nu,T_{d}) \Omega (1-e^{-\tau_{\nu}})
\end{equation}
where, $ S_{\nu}(\nu) $ is the observed flux density, $I_{\rm bkg}(\nu) $ is the background flux 
which in our case is obtained from the Gaussian fit, $ B_{\nu}(\nu,T_{d}) $ is the Planck's function, $T _{d} $ is the dust 
temperature, $ \Omega $ is the solid angle (in steradians) from where the flux is obtained (solid angle subtended by a 
14$'' \times $ 14$''$ pixel) and $\tau_{\nu} $ is the optical depth. The optical depth in turn is given by,
\begin{equation}
\tau_{\nu} = \mu_{\rm H_{2}} m_{\rm H} \kappa_{\nu} N(\rm H_{2}) 
\end{equation}
where, $\mu_{\rm H_{2}}$ is the mean molecular weight, $ m_{\rm H}$ is the mass of hydrogen atom,
$ \kappa_{\nu}$ is the dust opacity and $N(\rm H _{2} $) is the column density. 
We assume a value of 2.8 for  $\mu_{\rm H_{2}}$ \citep{2008A&A...487..993K}. The dust opacity $ \kappa_{\nu} $ is defined 
to be $ 
\kappa_{\nu} = 0.1~(\nu/1000~{\rm GHz})^{\beta}~{\rm cm^{2}/g}$. $\beta$ is the dust emissivity spectral
index which is assumed to be 2 \citep{{1983QJRAS..24..267H}, 
{1990AJ.....99..924B}, {2010A&A...518L.102A}}. 
The non-linear least square Levenberg-Marquardt fitting algorithm
is used. We include 15\% uncertainty in the background subtracted Hi-GAL fluxes (Launhardt et al. 2013). 
Dust temperature and column density are taken as free parameters in the model. From the fitted 
values, the temperature and column density maps are generated. 

According to \citet{2010ApJ...724L..44C}, the contribution from the stochastically heated very small grains (radius $ < 
$~0.005 $ \mu $m) 
to the PACS 70 $ \mu $m channel can be as large as $ \sim $50\%. Hence, a single modified 
blackbody model fit would possibly overestimate the cold dust temperatures and a two-temperature gray body 
is therefore essential to represent the emission from 70~$\mu $m \citep{2012MNRAS.425..763G}. 
Several studies have excluded the 
70~$\rm \mu m$ \citep{{2011A&A...535A.128B},{2012A&A...540A..10S}} while generating the above maps. 
However, we present maps with and without 70~$\rm \mu m$ emission and investigate the effect of this waveband on the 
modeled column density and dust temperature distribution. 

Figure \ref{temp_cdens} shows the column density and dust temperature distribution of the region associated
with IRAS 20286+4105. In the figure we have maintained same scaling for images with and without 70~$\rm \mu m$ 
emission for ease of comparison. 
The morphology is similar in both the column density maps with a single clump which traces the IRAS 20286+4105 cloud. 
The densest part of the cloud is seen to be coincident with the position of the IRAS point source. 
However, the pixels in the map generated without 70~$\rm \mu m$ flux fit to higher values. The median of the
fitted column density values are $\rm 1.6\times10^{21}\,cm^{-2}$ and $\rm 2.9\times10^{21}\,cm^{-2}$ with
and without 70~$\rm \mu m$ flux, respectively. In comparison, the 
morphology changes in the temperature maps where the one generated without 
70~$\rm \mu m$ emission displays a relatively elongated distribution. Here, the pixels in the map generated by 
including the 70~$\rm \mu m$ fits to higher temperatures as expected and discussed earlier. The median of the
fitted dust temperature values are 24.6~K and 19.9~K with and without 70~$\rm \mu m$ data, respectively. 

\begin{figure*}
\centering
\includegraphics[scale=0.35]{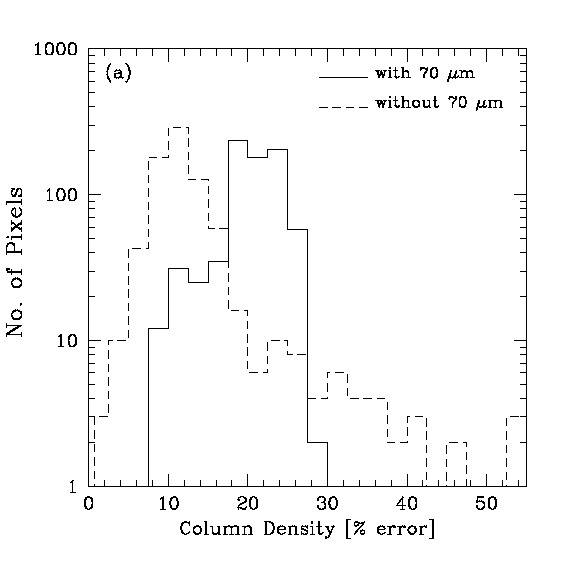}
\includegraphics[scale=0.35]{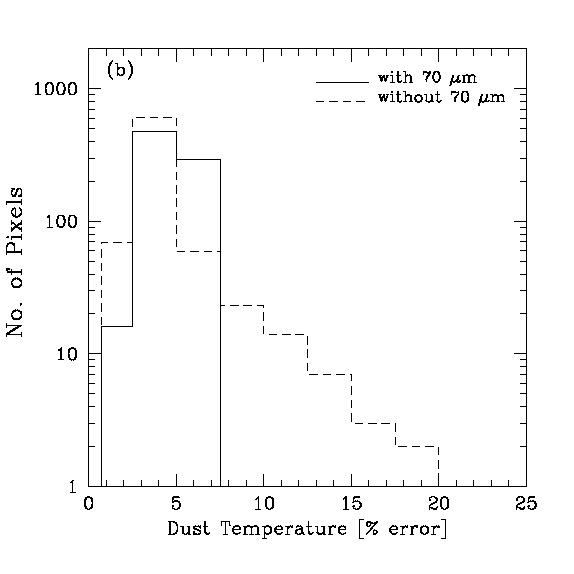}
\caption{Histograms pixel distribution of the percentage error in the fitted values for (a) Column density and (b) Dust 
temperature. 
The solid and dashed lines are including and excluding 70~$\rm \mu m$ emission, respectively.}
\label{temp_cdens_error}
\end{figure*}

For deriving the physical parameters, we inspect the fitting errors in both set of maps. 
In Figure \ref{temp_cdens_error}, we plot the pixel distribution of the percentage errors on the fitted values for the 
generated maps shown
in Figure \ref{temp_cdens}. As is evident, the percentage errors in the column density maps peak at lower
values if we exclude 70~$\rm \mu m$ flux with the median error in the derived column density being $\sim$ 20 
and 12\% with and without 70~$\rm \mu m$ data, respectively. However, the  percentage error reaches as high 
values as 55\% if we exclude 70~$\rm \mu m$ emission. 
In addition, all the pixels having fitting errors above 20\% lie inside the IRAS 20286+4105 cloud. This amounts
to half the pixels of the cloud. Whereas, the percentage of such `bad' pixels reduces to only 10\% if we include all five 
{\it Herschel} bands. The errors in both the temperature maps are below 
20\%, with the median error in the derived dust temperature being $\sim$ 5 
and 4\% with and without 70~$\rm \mu m$ flux, respectively. Hence, for mass estimation (see next section), we use the maps 
generated including the 70~$\rm \mu m$ data point.

\subsubsection{Properties of dust clumps}
\label{clump_prop}
\begin{figure}
\centering
\includegraphics[scale=0.38]{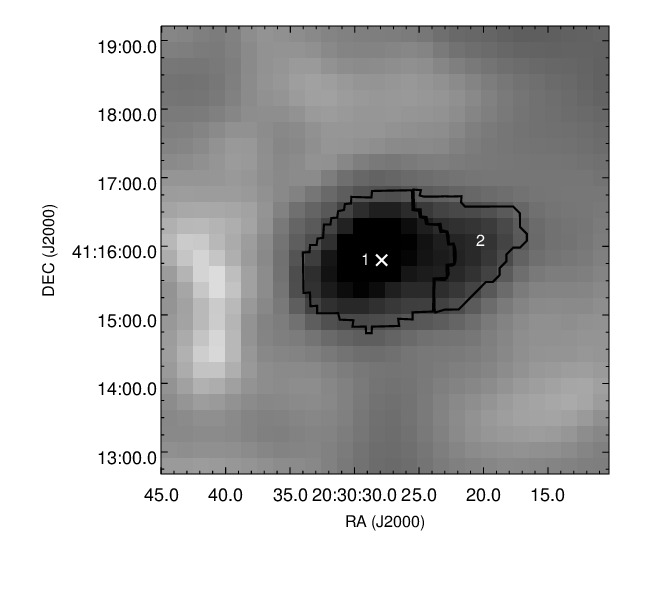}
\caption{ Clump apertures identified using \textit{clumpfind} algorithm marked over the derived column density map
shown in Figure \ref{temp_cdens}.}
\label{clump}
\end{figure}

The low resolution of the column density maps restrain us in resolving sub-clumps, if any, within this central clump. 
Hence, we use the 250~$\rm \mu m$ image and the 2D variation of the \textit{clumpfind} algorithm 
\citep{2011ascl.soft07014W} to detect and identify sub-clumps. The  250~$\rm \mu m$ image has a higher and
optimum resolution of 18\arcsec.
A threshold contour level of 630 MJy/Sr (which corresponds to $\rm 18\sigma$) is used
with a contour spacing of $\rm 3\sigma$. The rms noise of the map, $\sigma$, is estimated to be 35 MJy/Sr . 
The two sub-clumps identified using the 250~$\rm \mu m$ image are shown in Figure \ref{clump}.

Masses of these two clumps associated with IRAS 20286+4105 are estimated from the derived column density map using the 
following expression from \citet{2012A&A...547A..11N}.
\begin{equation}
\rm M_{clump} = \mu_{\rm H_{2}} m_{\rm H} A_{\rm pixel} \Sigma {\it N} (\rm H_{2} ) 
\label{clump_mass1}
\end{equation}
where m$ _{\rm H} $ is the mass of hydrogen, A$ _{\rm pixel} $ is the pixel area in cm$ ^{2} $, $ \mu_{\rm H_{2}} $ is the 
mean molecular weight and $\rm  \Sigma {\it N} (H_{2} ) $ is the integrated column density over the pixel 
area. The clump apertures retrieved from the {\it clumpfind} algorithm are used to obtain the integrated column density. 
Masses of Clump 1 and 2 are estimated to be $\sim$ 173 and 30~$\rm M_{\odot}$, respectively. 
                                                                                       
We also determine the masses of the sub-clumps from the 250~$\rm \mu m$ image using the 
following expression from \citet{2008A&A...487..993K},
\begin{equation}
\begin{aligned}
\rm M = 0.12 ~\rm M_{\odot} \left(e^{1.439(\lambda /{\rm mm})^{-1}({\it T}_{\it d}/10 K)^{-1}}-1 \right) \\ \times \left(\frac{{\kappa}_{\nu}}{\rm 0.01 cm^{2} g^{-1}} \right)^{-1} \left( \frac{S_{\nu}}{\rm Jy} \right) \left(\frac{D}{\rm 100 pc} \right)^{2} \left( \frac{\lambda}{\rm mm} \right)^{3}
\end{aligned}
\label{clump_mass2}
\end{equation}
\medskip
where $T _{d} $ is the dust temperature, $ {\kappa}_{\nu} $ is the dust opacity which is 
taken as  $0.1 (\frac{\nu}{\rm 1000 GHz})^{\beta}$, {\it D} is the distance, $ S_{\nu} $ is the integrated 
flux. $T _{d}$ is assumed to be the median value of the dust temperature within the aperture of the clumps.
The derived masses and other physical properties of the clumps are listed in Table \ref{clump_table}.
The volume densities, $n_{\rm H_2}$, are estimated assuming spherical clumps of uniform density.
Masses derived using the column density maps are on the lower side compared to those derived from the 
250~$\rm \mu m$ fluxes. We adopt the former values since these are determined using
all five {\it Herschel} images and hence would be a better estimate. The total mass of the clumps associated with IRAS 
20286+4105 is 203~M$_{\odot}$ which
is higher as compared to the estimate of 160~M$_{\odot} $ obtained by \citet{2011ApJ...727..114R} using 
data from the BLAST survey. It should be noted here that the effective radius used by them is lower (0.5~pc) 
compared to our combined clump size.    

\begin{table*}
\caption{Physical properties of the clumps.}
\begin{footnotesize}
\begin{tabular}{cccccccc}
\hline
\\
Clump No & Dust Temp.  & Effective Radius  & $n_{\rm H_2}$ & Clump Mass  & Clump Mass \\  
         &             &                   & &(from 250~$\rm \mu m$) & (from column density map) \\
         & (K) & (pc)& ($\rm \times 10^3 cm^{-3}$) & (M$ _{\odot}$) & (M$ _{\odot}) $  \\
\hline
\\
1      & 25.7   & 0.47 & 5.8 & 197  & 173 \\
2     & 25.8   & 0.34 & 2.7 & 40   & 30 \\  
\\
\hline
\end{tabular}
\end{footnotesize}
\label{clump_table}
\end{table*}

Clump 1 has tracers of ongoing star formation. It
is seen to be associated with intermediate-mass stars, masers, 24~$\rm \mu m$ emission from warm dust and outflows. 
All the six identified candidate YSOs are located within Clump 1. Except an arc of faint 24~$\rm \mu m$ emission,
Clump 2 does not show any signature of active star formation region and hence seems to be a potential new site for future generation of star formation.  

\subsubsection{Variation in Dust Emissivity Index}
\label{beta_td}

\begin{figure*}
\centering
\includegraphics[scale=0.35]{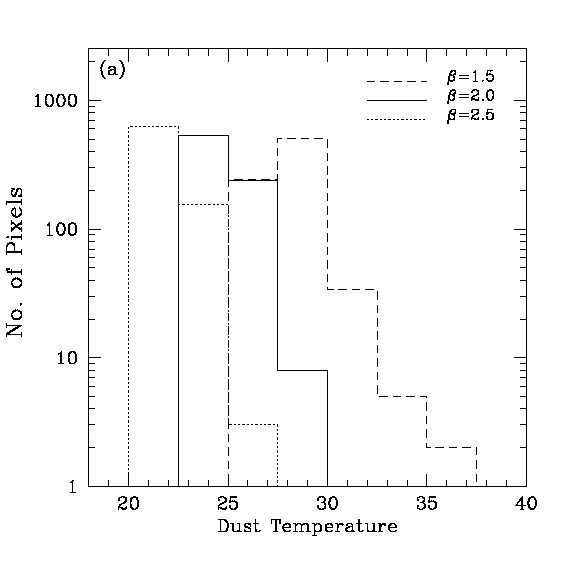}
\includegraphics[scale=0.35]{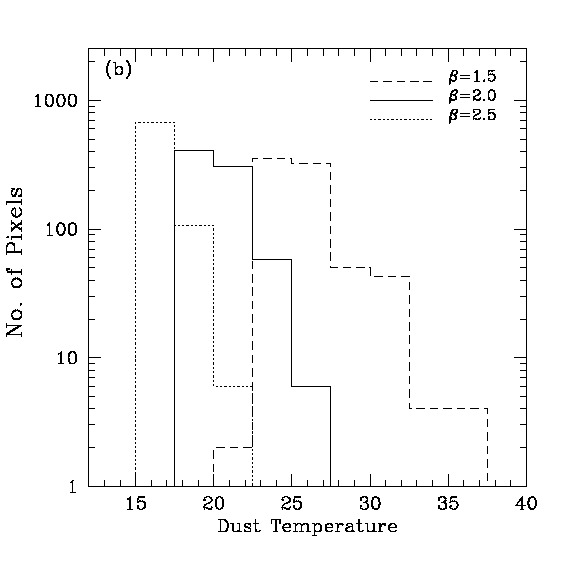}
\caption{Histogram of pixels values for dust temperature from maps generated with three fixed values of 
$\beta = 1.5, 2.0, 2.5$ (a) including 70~$\rm \mu m$  
flux and (b) excluding 70~$\rm \mu m$ flux.} 
\label{beta_hist}
\end{figure*}

There is growing evidence of an inverse relationship between the dust emissivity
spectral index, $\beta$ and the dust temperature, $T _{d} $ \citep{{2003A&A...404L..11D}, {2008A&A...481..411D}, 
{2010A&A...520L...8P}, {2010ApJ...713..959V}, {2012A&A...542A..10A}}.
We investigate this for the region associated with IRAS 20286+4105. 
Temperature maps are generated for two additional values of 
$ \beta $ = 1.5 and 2.5. The histograms of the dust temperature maps are shown in Figure \ref{beta_hist}.
Here, pixels having error less than 20\% on the fitted temperature values are chosen.  
From the histograms, an inverse relation of $\beta$ with the dust temperature is evident. The peak of the
distribution shifts to lower temperatures as $\beta$ increases. 

\begin{figure*}
\centering
\includegraphics[scale=0.42,bb=30 0 536 392,clip]{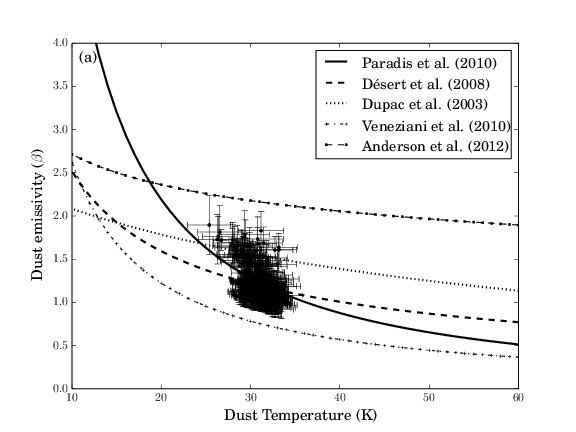}
\includegraphics[scale=0.42,bb=30 0 536 392,clip]{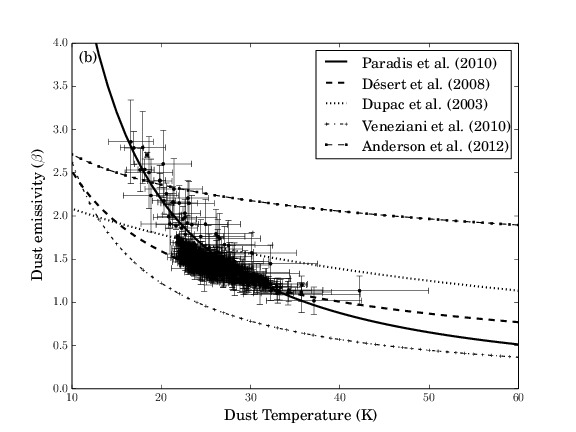}
\caption{Variation of $\beta$ with dust temperature for a $\rm 5\arcmin \times 5\arcmin~$ 
region centered on the IRAS point source. (a) Plot from maps including 70~$\rm \mu m$ flux and
(b) Plot from maps excluding 70~$\rm \mu m$ emission.
Fits from literature \citep{{2003A&A...404L..11D}, 
{2008A&A...481..411D}, {2010A&A...520L...8P}, {2010ApJ...713..959V}, {2012A&A...542A..10A}} 
are shown.}
\label{beta_T_plot}
\end{figure*}

Next we keep $ \beta$ as a free parameter along with temperature and column density in the model. 
We proceed further for a pixel wise analysis of the generated $\beta$ and dust temperature maps. 
This enables us to probe the correlation between these two parameters in the vicinity of a 
star forming region at a resolution of 36$''$. We select only
those pixels from our $\beta$ and $T_{d}$ maps which have errors less
than 20\% in both parameter space. This procedure is inspired from the work
of \citet{2003A&A...404L..11D}, who explain the existence of a degeneracy between these two
quantities which is due to the inherent large error bars on $\beta$ for 
colder regions and large errors for temperature for warmer components. Figure \ref{beta_T_plot} plots 
the pixel values of $\beta$ as a function of the dust temperature for a $\rm 5\arcmin \times 5\arcmin~$ 
region centered on the IRAS point source. 
It should be noted here that in the maps generated, without the 70~$\rm \mu m$ data, 
a single pixel was found with errors in both parameters less than 5\% but a high fitted value of 
temperature ($\sim 69$~K) and a low value of $\beta$ ($\sim 0.7$).    
In Figure \ref{beta_T_plot}, we also show fits from other studies showing the relation between
these two parameters. 
The values of $\beta$ and $T_{d}$ obtained 
from the maps generated including the 70~$\rm \mu m$ emission hints at the inverse relation whereas, the
values obtained without including the 70~$\rm \mu m$ emission clearly shows an inverse correlation
between the dust emissivity index and the temperature. As discussed earlier, including 70~$\rm \mu m$ emission
possibly overestimates the cold dust temperature and this could be the reason for the above difference seen. 
The result obtained in the latter case is 
consistent with the best fit curve estimated by \citet{2010A&A...520L...8P}. The lower temperature region
is also well correlated with the above fit. Their results are based on two Hi-GAL fields at Galactic
longitudes, $l = 30^{\circ}$ and $l = 59^{\circ}$. 
There is a high density of points in the temperature range 
$\sim 22 - 35$~K, where a significant number of pixels seem to 
follow the fit proposed by \citet{2008A&A...481..411D}.
Their results are based on submillimetre point
sources from the Archeops experiment. As discussed in \citet{2015arXiv151200987V} and references
therein, this anti-correlation is attributed to the intrinsic properties of the grains in the cold
dust environments. The warmer regions can be understood as comprising of low emissivity index bare silicate 
aggregates or porous graphite grains \citep{{2003A&A...404L..11D},{1989ApJ...341..808M}}, while the
regions with higher $\beta$ values could have the composition of icy mantles \citep{1975ApJ...200...30A}.

\subsection{General star formation scenario in IRAS 20286+4105}
The `runaway' star hypothesis suggests that the massive B0 - B0.5 star responsible for the
ionized emission is not formed from the IRAS 20286+4105 cloud but instead has its birthplace
in the supernova remnant. The ongoing star formation activity in this cloud thus includes
the six Class I sources seen clustered towards the cloud centre. The SED modelling of these
indicate that five of them are intermediate mass stars. In the absence of massive stars, the bright 24~$\rm \mu m$ MIR emission is mostly due to this cluster of intermediate-mass Class I sources. This however does not preclude the
presence of further deeply embedded young protostars contributing towards heating the dust
which radiates at 24~$\rm \mu m$. The total luminosity of the six sources add up to 
$\rm \sim 10^3 L_{\odot}$ which is an appreciable fraction of the total luminosity of
5.3 $ \times 10^{3} $ L$ _{\odot}$ cited earlier. Thus, there seems to be no conclusive
evidence of high-mass star formation in this cloud.

The above discussion finds support in the nature of Clump 1 which hosts the embedded
sources. For the estimated radius of 0.5~pc, the mass should be greater than $\rm \sim 320\, M_{\odot}$ for the clump to form high-mass stars \citep{2010ApJ...723L...7K}. This is
almost twice the value estimated from the column density map. Hence, following the empirical
mass-radius relation given by \citet{2010ApJ...723L...7K}, this clump should be devoid 
of massive stars. So is the case with Clump 2 for which the threshold mass is calculated to be
$\rm \sim 210\, M_{\odot}$ which is almost seven times higher than that estimated from
the {\it Herschel} maps. Another criterion for clumps to be high-mass star forming sites was
proposed by \citet{2008Natur.451.1082K}, who predicted a threshold surface density ($\Sigma_{th}$) of $\rm \sim 1\, g\, cm^{-2}$ for massive stars to form as opposed to fragmentation into lower masses. The estimated values of $\Sigma$ for Clumps 1 and 2 are 0.05 and 0.02  $\rm g\, cm^{-2}$, 
respectively. The estimate of Clump 1 is marginally lower than the value of 
0.117 $\rm g\, cm^{-2}$ determined by \citet{2011ApJ...727..114R}. 
These estimated $\Sigma$ values are far below $\Sigma_{th}$ = 1. In a detailed study of
massive star forming cores, \citet{2009ApJS..181..360C} have shown that only for a small fraction 
cores $\Sigma > \Sigma_{th}$ and the median values of both active and quiescent cores 
are $\rm \sim 0.2\, g\, cm^{-2}$. As discussed by these authors, the low values of $\Sigma$
could be attributed to the large clump apertures adopted and the actual core mass densities
could be higher and possibly closer to $\Sigma_{th}$. However, a point worth noting here is
that, for both the clumps associated with IRAS 20286+4105, the $\Sigma$ values are $\sim 5 - 13$
lower than the values estimated for active and quiescent cores by \citet{2009ApJS..181..360C}. 
Thus, given the mass threshold and the surface mass density estimates, the picture of a runaway star and triggered population of intermediate-mass Class I objects augurs well with the physical condition of the clumps which do not qualify as massive star forming sites.

\section{Conclusion}
\label{conclusion}
In this paper we carried out an extensive multi-wavelength study of the star forming
region IRAS 20286+4105 and its associated environment. Our main inferences are as follows

\begin{enumerate}

\item Using UKIDSS NIR data, we show the effect of sensitivity
on the detection of sparsely populated clusters. The 2MASS cluster known to be associated
with IRAS 20286+4105 is not detected with the deeper UKIDSS data. This could possibly 
be due to the overwhelming background sampled which suppresses this low stellar density
cluster/group.

\item As deduced from the MIR and NIR data, the region is an active star forming complex harboring a cluster of 
six Class I YSOs towards the center of the cloud. SED modeling indicates the YSOs to be mostly intermediate mass stars. Several detected masers are seen to be located in the close vicinity of these Class I sources. 

\item  GMRT radio continuum observations at 610 and 1280~MHz in conjunction with VLA 4.89~GHz
archival data show the presence of a cometary HII region consistent with the morphology
obtained for ionized gas in optical (H$\alpha$) and NIR (Br$\gamma$). Assuming optically
thin free-free emission, the spectral type of a single ionizing star responsible for this
is estimated to be B0.5 - B0.

\item The bow-shock model seems to be the likely mechanism responsible for the cometary radio morphology. The projected  
stand-off distances at 1.61~kpc are estimated to lie between $\sim$ 2\arcsec - 13\arcsec, which indicate that the ionizing source
lies within the radio arc. If we assume the radio peak to be associated with the ionizing source then the
stand-off distance is consistent with the distance of the edge of the radio
arc from the radio peak position. The above argument of the bow-shock model finds support in 
the morphological indication of a possible association with a supernova remnant SNR G079.8+1.2 
which might be responsible for the supersonic motion of the ionizing star. Kinematic age 
estimate of the runaway star responsible for the bow-shock supports the binary 
supernova scenario of high-velocity stars.

\item The Class I sources located inside the cloud could be a population formed due
to the collapse of a pre-existing clump triggered by an encroaching shock associated with the
supernova. Further evidence of triggered star formation is seen as an enhanced density of
YSOs towards the east and bordering the radio arc.

\item Apart from the diffuse emission, we detect a radio knot (RS1) which shows a negative
spectral index ($\rm \sim -0.5\pm0.05$). This knot is seen to be coincident with
the position of an outflow and hence is likely to be due to the non-thermal emission associated
with it. 

\item From the FIR {\it Herschel} data, we generate the dust temperature and column density maps.
The presence of a central dust clump is clearly seen in the column density map. Using the 250~$\rm \mu m$
and the 2D variation of the {\it clumpfind} algorithm, this central clump is resolved into two sub-clumps
with estimated masses from the column density maps to be 173$\,{\rm M}_{\odot}$ (Clump 1) and 30$\,{\rm M}_{\odot}$ (Clump 2).
 
\item Clump 1 is seen to be an active star forming region, whereas
Clump 2 shows no sign of ongoing star formation.

\item The mass, radius and surface density of the clumps indicate that they are
likely to be devoid of massive star formation.

\item We investigate the relation between the dust temperature and the dust
emissivity spectral index in the vicinity of the IRAS 20286+4105
star forming complex. Variation is seen in the spectral index which shows an 
anti-correlation with dust temperature.
\end{enumerate}

\begin{footnotesize}
Acknowledgment : We would like to thank the referee for his/her valuable suggestions which helped
in improving the quality of the paper. VR would like to thank Dr. Ishwara Chandra of NCRA, Pune for organizing a workshop for introduction to radio data analysis.
We thank the staff of the GMRT, that made the radio observations possible. GMRT is run by the National Centre for Radio Astrophysics of the Tata Institute of Fundamental Research. 
We thank the staff of IAO, Hanle and CREST, Hosakote, that made the NIR observations possible. The facilities at IAO and CREST are operated by the Indian Institute of Astrophysics, Bangalore.
\end{footnotesize}

\bibliography{mybib}
\end{document}